\documentclass[%
 amsmath,amssymb,
 aps,
showkeys
]{revtex4-2}

\usepackage{graphicx}
\usepackage{dcolumn}
\usepackage{bm}
\usepackage{color}

\usepackage[margin=1.25in]{geometry}

\graphicspath{{./figures/}}

\newcommand{\etaVE}{\eta_{\text{ve}}}  
\newcommand{\etaV}{\eta_{\text{v}}}    

\newcommand{\RE}{\mathrm{Re}}  
\newcommand{\IM}{\mathrm{Im}}  
\newcommand{\DE}{\mathrm{De}}  

\renewcommand{\c}{c} 
\newcommand{\sigmacrit}{\sigma_{0}}

\newcommand{\X}{\mathbf{X}}
\newcommand{\Fp}{\mathbf{F}^{p}}
\newcommand{\F}{\mathbf{F}}
\newcommand{\C}{\mathbf{C}}

\newcommand{\Ffluid}{\mathbf{F}^{\text{fluid}}}  
\newcommand{\Fsol}{\mathbf{F}^{\text{sol}}}      
\newcommand{\Fpol}{\mathbf{F}^{\text{pol}}}      
\newcommand{\fpol}{f^{\text{pol}}}               
\newcommand{\fpolhat}{\hat{f}^{\text{pol}}}      

\newcommand{\singlepanelwidth}{0.60\textwidth}
\newcommand{\singlepanelwidthwide}{0.75\textwidth}

\definecolor{review_color}{rgb}{0.0, 0.0, 0.0}  
\newcommand{\change}[1]{\textcolor{review_color}{#1}}

\begin{document}

\preprint{APS/123-QED}
\title{Effect of fluid elasticity on the emergence of oscillations in an active elastic filament}
%


\author{Kathryn G.\ Link}
\affiliation{%
 Department of Mathematics, University of California, Davis, Davis, CA 95616
}%
\author{Robert D.\ Guy}
\affiliation{%
 Department of Mathematics, University of California, Davis, Davis, CA 95616
}%
\author{Becca Thomases}
\affiliation{%
 Department of Mathematical Sciences, Smith College, Northampton, MA, 01063
}%

\author{Paulo E.\ Arratia}%
\affiliation{
Department of Mechanical Engineering and Applied Mechanics,
The University of Pennsylvania,  Philadelphia, PA 19104
 }%


\date{\today}

\begin{abstract}

%
%
%

Many microorganisms propel  through complex media by deformations of their flagella. The  beat is thought to emerge from  interactions between  forces of the surrounding fluid, passive elastic response from deformations of the flagellum, and active forces from internal molecular motors. The beat varies in response to changes in the fluid rheology, including elasticity, but there is limited data on how systematic changes in elasticity alters the beat. This work analyzes a related problem with fixed-strength driving force: the emergence of beating of an elastic planar filament driven by a follower force at the tip in a viscoelastic fluid. This analysis examines how the onset of oscillations depends on the strength of the force and viscoelastic parameters. Compared to a Newtonian fluid, it takes more force to induce the instability in viscoelastic fluids, and the frequency of the  oscillation is higher. The linear analysis predicts that the frequency increases with the fluid relaxation time. Using numerical simulations, the model predictions are compared with experimental data on frequency changes in bi-flagellated alga \textit{Chlamydomonas reinhardtii}. The model shows the same trends in response to changes in both fluid viscosity and Deborah number, and thus provides a \change{possible} mechanistic understanding of the experimental observations.

\end{abstract}


\maketitle


\clearpage

%
%
\section{Introduction}
\label{sec:introduction} 

Many microorganisms, such as sperm, propel themselves through complex media by deformations of their flagella.  It has long been observed that the rheology of the surrounding fluid alters the shape and frequency of the flagellum beat of mammalian sperm \cite{katz1978movement,Suarez:1992:BR:hyperactivation,Ishijima:1986:GR:sperm_shape,Smith:2009:CM:humanspermviscosity,Guasto:2020:RSI:flagella_shape}, of sea urchin and related marine animal sperm \cite{Brokaw:1966:JEB:Viscosity,Woolley:2001:JEB:helicalandplanar}, and of the bi-flagellated alga \textit{Chlamydomonas reinhardtii} \cite{qin2015flagellar,geyer2022ciliary}. In addition to beat changes, fluid viscosity has been show to affect the coordination in arrays of cilia that drive cell locomotion \cite{machemer1972ciliary} and transport mucus \cite{Gheber:1998:CM:viscosity_cilia_mucus}. 

Human sperm and sperm from marine invertebrates exhibit different gait changes in response to high viscosity environments \cite{Guasto:2020:RSI:flagella_shape}. It has been hypothesized that the gait changes in mammalian sperm are important for fertilization because they must swim through viscoelastic mucus \cite{Suarez:1992:BR:hyperactivation,Guasto:2020:RSI:flagella_shape}. There have been multiple observations of sperm gaits in mucus and other viscoelatic fluids \cite{katz1978movement,Suarez:1992:BR:hyperactivation,Ishijima:1986:GR:sperm_shape,Smith:2009:CM:humanspermviscosity}, but there has been no systematic documentation of how gradual changes in fluid elasticity affect the gait. Changes in the beat frequency, shape, and swimming speed of \textit{Chlamydomonas reinhardtii} in response to both viscous and elastic properties of the surrounding fluid were recently documented \cite{qin2015flagellar}; it was observed that the beat frequency was enhanced by fluid elasticity, and the frequency changed nonmonotonically with fluid viscosity in viscoelastic fluids. 

Fluid elasticity clearly influences the flagellum beat, but the physical mechanism for how fluid rheology shapes the beat is not known. There have been theoretical studies of how fluid elasticity affects the gait for prescribed active motor forces \cite{fu2008beating,thomases2017role}. These studies considered the active forces as a traveling wave with a given frequency, and they examined how the fluid rhelology affected the resulting shapes. This approach was able to explain the qualitative shape change observed in some sperm species in viscoelastic fluids \cite{Ishijima:1986:GR:sperm_shape}, and the increased amplitude of the beat, and thus increased swimming speed, in artificial swimmers with flexible tails \cite{Espinosa-Garcia:2013:POF:flexible}. However because the frequency of the active forces was prescribed, this approach cannot be used to understand how the beat frequency changes with fluid rheology as observed in \cite{qin2015flagellar}.

The flagellum beat is powered by dynein motors that form crossbridges between the microtubule doublets that make up the axoneme. These motors generate active shear forces between adjacent doublets that through interactions with other passive forces and constraints at the base leads to bending \cite{satir1968studies}. It is not understood how the motors along the flagellum are coordinated spatially and temporally to produce the observed waves of bending. There are different hypotheses about how mechanical feedback on motor activity from deformations of the flagellum lead to emergent coordination of the whole system. Some of the leading feedback mechanisms that have been explored are that the motors respond to changes in curvature \cite{Brokaw:1971:JEB:bendpropgation,BROKAW:1972:BJ:computersimulation,hines1978bend}, tangential deformations (i.e.\ sliding control) \cite{Camalet_2000,Riedel-Kruse:2007:HSFP:motorsshape}, or normal forces (i.e.\ ``geometric clutch") \cite{LINDEMANN:1994:JTB:GC,BAYLY:2014:BJ:doubletseparation}. All of these mechanisms have been explored thoroughly in models, and they are all capable of producing emergent waves in the flagellum. Analysis of the bifurcation structure of the three models cast doubt on the sliding-control mechanism \cite{bayly2015analysis}, though sliding control was capable in matching experimental data on bull sperm \cite{Riedel-Kruse:2007:HSFP:motorsshape}. A comparison of all three models on data on \textit{Chlamydomonas reinhardtii} data favors curvature control \cite{Sartori:2016:Elife:dynamiccurvature}. Despite years of theoretical effort, it is not clear if any of these feedback mechanisms are involved in producing the flagellum beat.

%
In \cite{bayly2016steady} an alternative mechanism was proposed and analyzed for producing the flagellum beat that does not require dynein regulation nor spatiotemporal organization of dynein activity. The mechanism suggested in \cite{bayly2016steady} is related to a dynamic instability known as flutter that results when an elastic structure in fluid is subject to axial loading. Dynein generates a tension that buckles the filament. Unlike static buckling, the direction of the motor forces remains tangent to the filament as it deforms which results in an oscillatory motion.

%
The instability analyzed in \cite{bayly2016steady} is similar to the instability of filaments under external load in the tangent direction known as a ''follower force" (because it follows the direction of the filament) \cite{Herman:1964:JAM:stabilitynonconservative}. There have been several recent analyses of filaments subject to a follower force at low Reynolds number inspired by the motion of biological filaments driven by molecular motors \cite{Sekimoto:1995:PRL:symmetrybreaking,de2017spontaneous,Feng:2018:RSI:instabilityoscillations,Fily:2020:RSI:bucklinginstabilities,Stein:PRL:2021:swirling,PhysRevLett.125.148101,PhysRevFluids.6.L121101}. These models have been used to understand observations in \textit{in vitro} motility assays \cite{Sekimoto:1995:PRL:symmetrybreaking}, cytoplasmic streaming \cite{Stein:PRL:2021:swirling}, beating flagella and cilia \cite{Feng:2018:RSI:instabilityoscillations}, \change{and coordination between pairs \cite{PhysRevLett.125.148101} and arrays of cilia \cite{PhysRevFluids.6.L121101}}. These works have thoroughly analyzed the Hopf bifurcation from rest to a beating pattern including how different boundary conditions and restrictions (i.e.\ planar vs.\ 3D) result in different dynamics in viscous fluids \cite{de2017spontaneous,Feng:2018:RSI:instabilityoscillations,Fily:2020:RSI:bucklinginstabilities}.

In this paper we analyze the emergence of oscillations of a planar elastic filament pinned at the base subject to a follower force at the tip in a viscoelastic fluid. We examine how the elasticity of the surrounding fluid affects the strength of the applied force needed to produce the oscillation and the emergent frequency at the bifurcation. Our results show that the critical value of the force at which oscillations occur is greater in a viscoelastic fluid than in a Newtonian fluid, and the frequency of the beat is always increased by the elasticity of the fluid. We compare the model predictions for how the frequency changes with relaxation time and total viscosity with the experimental measurements from \cite{qin2015flagellar}. Our analysis captures the observed frequency increases with relaxation time and the nonmonotonic frequency response to changes in viscosity, and thus offers a possible mechanistic explanation for how the beat frequency is affected by fluid elasticity.

%
%
\section{Model and Equations}\label{Sec:Model}

\begin{figure}[b]
  \centering
  \includegraphics[width = \singlepanelwidth]{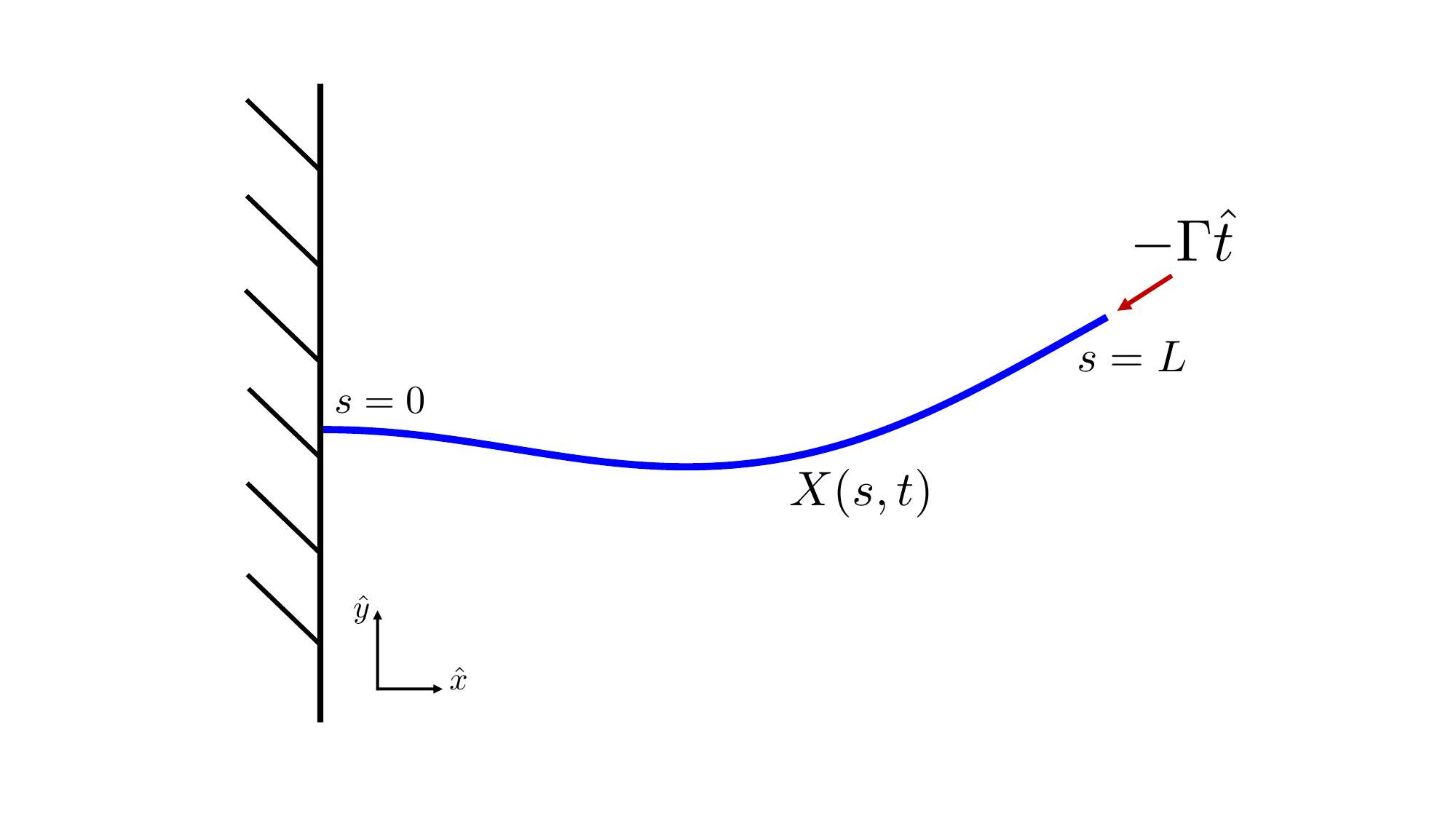}
 \caption{Schematic of a horizontal flexible filament clamped at one end with a follower force of strength $\Gamma$ applied at its tip. The filament position is defined as $\X(s,t)$,  where $0 \le s \le L$ is the arc length coordinate. The local tangent vector is $\X_{s}=\hat{{\bf{t}}}(s,t)$.}
 \label{fig:schematic}
 \end{figure}

We consider the motion of a slender, inextensible, planar, elastic filament, clamped at one end and subject to a compressive follower force of strength $\Gamma$ at the tip of the filament in a viscoelastic fluid at zero Reynolds number. The mathematical model is analogous to that presented in \cite{de2017spontaneous} with the addition of fluid viscoelasticity. Let $0 \le s \le L$ be the arclength where $s=0$ corresponds to the clamped base. The position of the filament is $\X(s,t)=(x(s,t),y(s,t))$; see Figure \ref{fig:schematic}.

The instantaneous force balance for the filament is 
\begin{equation}
 -k_{b}{\bf{X}}_{ssss} - (T {\bf{X}}_{s})_{s} + \Ffluid = 0, \label{eq:force_bal}
\end{equation}
where the first term is the force per unit length from bending, the second term represents the tension that enforces the inextensibility constraint $\lvert\X_{s}\rvert=1$, and $\Ffluid$ is the drag force from the surrounding fluid. 
We assume that the fluid drag can be expressed as the drag in a viscous fluid plus a drag accounting for the viscoelastic effects. One can think of the fluid as composed of a Newtonian solvent with the addition of polymers which are responsible for viscoelastic stresses.  Thus, the drag force is
\begin{equation}
 \Ffluid= \Fsol + \Fpol,
\end{equation}
where $\Fsol$ and $\Fpol$ are the drag forces due to the solvent and polymers, respectively. 

At the clamped end, we have the boundary conditions 
\begin{equation}
{\bf{X}}(0,t) = 0, \ \ \ \ \ \text{ and }  \ \ \ \ \ {\bf{X}}_{s}(0,t) = \hat{\bf{e}}_{x}, \label{eq:clamp_bc}
\end{equation}
while at the free end
\begin{align}
&{\bf{X}}_{ss}(L,t) = 0, \label{eq:free_bc1}\\
&-k_{b}{\bf{X}}_{sss}(L,t) - T(L,t){\bf{X}}_{s}(L,t) = -\Gamma {\bf{X}}_{s}(L,t),\label{eq:free_bc2}
\end{align}
which capture the fact that the filament is torque-free and that the force at the tail and the external force must balance. The compressive follower force, $\Gamma{\bf{X}}_{s}$, is applied tangentially to the tip of the filament. This non-conservative force drives the motion of the filament.

The viscous drag force due to the solvent $\Fsol$ acting on the filament from the surrounding flow is given by resistive force theory \cite{Gray_Handcock:JEB:1955:RFT} which provides a local relation between the local filament velocity, ${\bf{X}}_{t}$, and the hydrodynamic force per unit length. The viscous drag force per unit length is defined as
\begin{equation}
\Fsol = -(\xi_{||}^{s}\hat{{\bf{t}}} \hat{{\bf{t}}} + \xi_{\perp}^{s}\hat{{\bf{n}}}\hat{{\bf{n}}})\cdot {\bf{X}}_{t}, \label{eq:vis_drag}
\end{equation}
where $\hat{{\bf{t}}}$ and $\hat{{\bf{n}}}$ are the local tangent and normal unit vectors. The drag coefficients in the perpendicular and parallel direction are $\xi_{\perp}^{s}$ and $\xi_{||}^{s}$ are proportional to the solvent viscosity $\mu_{s}$. For example, for a cylinder of radius $b$ and length $L$, $\change{\xi_{\perp}^{s} = \mu_{s}\alpha}$, where $\change{\alpha = 4\pi/[\ln(L/b) + 1/2]}$ and $\xi_{\perp}^{s}/\xi_{||}^{s} \rightarrow 2$ as $L/b \rightarrow \infty$ \cite{Gray_Handcock:JEB:1955:RFT}. 

We are interested in the effects of viscoelasticity at and near the bifurcation at which oscillations first emerge, and are thus small in amplitude. Given this, we utilize a linear viscoelastic model to describe the polymeric force $\Fpol$:
\begin{equation}
\tau \Fpol_{t} + \Fpol = -(\xi_{||}^{p}\hat{{\bf{t}}} \hat{{\bf{t}}} + \xi_{\perp}^{p}\hat{{\bf{n}}}\hat{{\bf{n}}})\cdot {\bf{X}}_{t}, \label{eq:poly_force}
\end{equation}
where $\change{\xi_{\perp}^{p} = \mu_{p}\alpha}$, $\mu_{p}$ is the polymeric viscosity, and $\tau$ is the fluid relaxation time. In the limit of zero fluid relaxation time ($\tau = 0$) or zero polymer viscosity, $(\mu_{p} = 0)$, the polymeric force $\Fpol = 0$ and we recover the viscous Newtonian fluid.
The assumption of a linear viscoelastic model for the drag force on a deforming filament at small amplitude  was utilized by previous authors to analyze the effect of viscoelasticity on shape changes of flagellum shapes and bending filaments \cite{fu2008beating,thomases2017role}. Further, this assumption was numerically validated in \cite{thomases2017role} by comparing this linear model with numerical simulations that involve the nonlinear viscoelastic stress.

Equations \eqref{eq:force_bal}, \eqref{eq:vis_drag}, and \eqref{eq:poly_force} are nondimensionalized by rescaling lengths by $L$, time by the viscous time scale $ L^{4}(\xi_{\perp}^{s} + \xi_{\perp}^{p}) / k_{b}$, tension (the Lagrangian multiplier) by $k_{b}/L^{2}$, and polymeric force by $k_{b}/L^{3}$, yielding
\begin{align}
&-\mathbf{X}_{ssss}-(\hat{T} \mathbf{X}_{s})_{s} - \beta(\mathcal{R}\hat{\textbf{t}}\hat{\textbf{t}} + \hat{\textbf{n}}\hat{\textbf{n}}){\bf{X}}_{t} + \Fpol = 0, \label{eq:model1}\\
& \lambda \Fpol_{t} + \Fpol = - (1-\beta)(\mathcal{R}\hat{\textbf{t}}\hat{\textbf{t}} + \hat{\textbf{n}}\hat{\textbf{n}}){\bf{X}}_{t}  \label{eq:model2}.
\end{align}
Here $\hat{T} = T L^{2} / k_{b}$ is the dimensionless tension, $\beta = \mu_{s} / (\mu_{s} + \mu_{p})$ is the viscosity ratio of the viscoelastic fluid, $\mathcal{R} = \xi_{||}^{s} / \xi_{\perp}^{s} = \xi_{||}^{p} / \xi_{\perp}^{p}$ is the ratio of tangential and normal drag coefficients, and $\lambda = \tau k_{b} /  L^{4}(\xi_{\perp}^{s} + \xi_{\perp}^{p})$ is the dimensionless relaxation time. 
\change{Note that the ratio of tangential and normal drag coefficients depends only on the aspect ratio of the filament, not on the viscosity, and thus has the same value for the solvent and polymerirc fluid.}

Similarly rescaling equations \eqref{eq:free_bc1} and \eqref{eq:free_bc2} yields the following dimensionless free boundary conditions 
\begin{align}
&{\bf{X}}_{ss}(1,t) = 0, \label{eq:bc2}\\
&-{\bf{X}}_{sss}(1,t) - \hat{T}(1,t){\bf{X}}_{s}(1,t) = -\sigma {\bf{X}}_{s}(1,t),\label{eq:bc3}
\end{align}
where the dimensionless ratio between the strength of the force at the tip and the elastic force is defined as
\begin{equation}
    \sigma = \Gamma L^{2} / k_{b}.
\end{equation} 
Since the force is compressive $(\Gamma > 0)$, $\sigma$ is always positive.

%
%
\section{Bifurcation Analysis in a Viscoelastic Fluid}
\label{Sec:LSA}

In a viscous fluid, the strength of the follower force, $\sigma$, is the only nondimensional parameter. As analyzed in \cite{de2017spontaneous}, there is a critical strength of the follower force, $\sigmacrit$, below which the straight filament at rest is stable. At  $\sigma=\sigmacrit$ there is a supercritical Hopf bifurcation so that for $\sigma>\sigmacrit$ the filament exhibits sustained oscillations. In a viscoelastic fluid there are three dimensionless parameters to consider: the follower force strength, $\sigma$, the relaxation time, $\lambda$, and the viscosity ratio, $\beta$. In this section we analyze how the critical follower strength and the frequency of the emergent oscillation in a viscoelastic fluid depend on $\lambda$ and $\beta$.

We consider small amplitude deviations from the rest state of a straight filament. In the small deformation regime, the tension to leading order is constant and therefore equal to the external force applied to the tip \cite{de2017spontaneous,Camalet_2000}, so that  $\hat{T}(s,t) = \sigma$. The leading order equations for the vertical displacement, $y(s,t)$, and vertical component of the polymer force, $\fpol(s,t)$, are 
\begin{gather}
 -\beta y_{t} - y_{ssss} - \sigma y_{ss} + \fpol = 0,\label{eq:lin_eq1}\\
  \lambda \fpol_{t} + \fpol = -(1-\beta)y_{t}\label{eq:lin_eq2}.
\end{gather}
Deformations in the horizontal direction and changes to the tension occur at higher order in deformation. 
The resulting boundary conditions are 
\begin{equation}
y(0,t) = y_{s}(0,t) = y_{ss}(1,t) = y_{sss}(1,t) = 0.\label{eq:lin_eq3}
\end{equation}

\subsection{Relationship to Stability in a Viscous Fluid}\label{VEevalsec}
We assume solutions of the form $y(s,t) = \hat{y}(s)e^{\etaVE t}$ and $\fpol(s,t) = \fpolhat(s)e^{\etaVE t}$ in equations \eqref{eq:lin_eq1}-\eqref{eq:lin_eq3}. \change{The real part of $\etaVE$ quantifies the growth (or decay if negative) rate of perturbations in viscoelastic fluid, and its imaginary part gives the frequency of oscillations.}
%
Eliminating $\fpolhat$ we obtain
\begin{equation}
  -\hat{y}_{ssss} - \sigma\hat{y}_{ss} = \left(\beta \etaVE +  \frac{(1-\beta)\etaVE} { (\lambda\etaVE + 1)}\right) \hat{y}.
  \label{eig_val1:eq}
\end{equation}
We let $\mathcal{L}=-\partial_{ssss}-\sigma\partial_{ss}$ denote the operator acting on the space of functions that satisfy boundary conditions \eqref{eq:lin_eq3}. We express \eqref{eig_val1:eq} as the eigenvalue problem
\begin{equation}
  \mathcal{L} \hat{y} = \etaV \hat{y},
  \label{vis_eig_prob:eq}
\end{equation}
where 
\begin{equation}
  \etaV = \frac{(1-\beta)\etaVE} {\lambda\etaVE  + 1} + \beta \etaVE 
  \label{eq:ve_eig_val}
\end{equation}
denotes the eigenvalues of $\mathcal{L}$, the real part of which quantifies the growth rate of perturbations in a viscous fluid. Consistent with this idea, note that when either $\beta=1$ or $\lambda=0$, $\etaVE=\etaV$ because the viscoelastic fluid reduces to a viscous fluid.

\subsubsection{Instability in a Viscous Fluid is Necessary for Instability in a Viscoelastic Fluid}
We first show that for a given strength of the follower force, the system is unstable in a viscoelastic fluid only if the system is unstable in a viscous fluid. Equation \eqref{eq:ve_eig_val} relating the eigenvalues of the follower problem in the viscoelastic fluid to those in a  viscous fluid can be expressed as
\begin{equation}
\etaV = \frac{(1-\beta)\lambda|\etaVE|^{2}}{\left|1+\lambda\etaVE\right|^2} 
      + \left(\frac{(1-\beta)}{\left|1+\lambda\etaVE\right|^2} + \beta\right)
        \etaVE.
\label{ve_eig_val_rearranged:eq}    
\end{equation}
The real parts of $\etaV$ and $\etaVE$ are related by 
\begin{equation}
  \RE(\etaV)=\change{\alpha_{0}}\left(\etaVE\right) + \change{\alpha_{1}}\left(\etaVE\right)\RE(\etaVE),
\label{re_ev:eq}
\end{equation} 
where $\change{\alpha_{0}}$ and $\change{\alpha_{1}}$ are real valued, nonnegative functions of $\etaVE$. Therefore, if $\RE(\etaVE)>0$, then $\RE(\etaV)>0$. This establishes that instability in a viscous fluid is a necessary condition for instability in a viscoelastic fluid.

\subsubsection{More Force Required for Instability in Viscoelastic}
Let $\hat{\sigma}$ denote the value of the follower force at which the rest state become unstable in the viscoelastic fluid. Because $\RE(\etaVE)=0$ when $\sigma=\hat{\sigma}$, $\RE(\etaV)=c_{0}\left(\etaVE\right)>0$ from \eqref{re_ev:eq}. Because $\RE(\etaV)<0$ for all $\sigma<\sigmacrit$, it follows that $\hat{\sigma}>\sigmacrit$. Therefore it follows that the follower force required for instability in the viscoelastic case is always larger than the force required in the viscous case.

\subsubsection{Higher Frequency in Viscoelastic Fluid}
We show that for the same follower force, the frequency of the oscillation in the viscoelastic fluid is always larger the frequency in the viscous fluid. From equation \eqref{ve_eig_val_rearranged:eq}, the imaginary parts of $\etaV$ and $\etaVE$ are related by 
\begin{equation}
\IM(\etaV) =  \left(\frac{(1-\beta)}{\left|1+\lambda\etaVE\right|^2} + \beta\right)\IM(\etaVE).
\label{im_ev:eq}
\end{equation}
Assume that $\RE(\etaVE)>0$ so that the rest state is unstable. If follows that $\left|1+\lambda\etaVE\right|^2> 1$, and thus from \eqref{im_ev:eq}
\begin{equation}
\IM(\etaV) < \IM(\etaVE).
\end{equation}
Because the emergent frequency of the oscillation near the bifurcation is approximately the imaginary part of the eigenvalue, we conclude that viscoelasticity increases the frequency of the oscillation.

%
%
%

\subsection{Numerical Calculation of Eigenvalues}
\begin{figure}
\centering
\includegraphics[width=\textwidth]{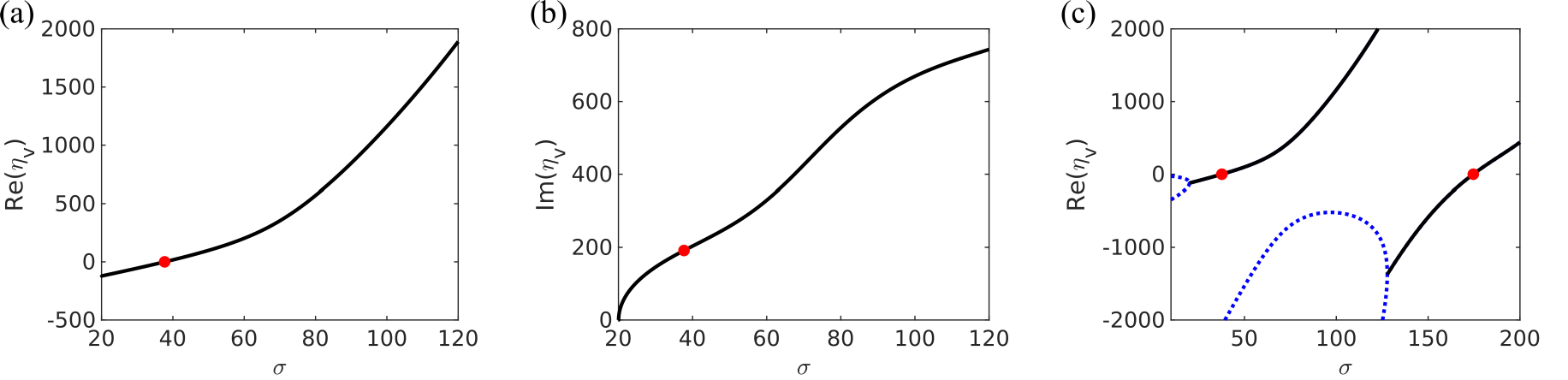}
\caption{Real part (a) and Imaginary part (b) of the eigenvalue with largest real part for a viscous fluid as a function of the follower force, $\sigma$. \change{(c) Real part of the four eigenvalues with largest real parts. Blue dashed lines denote real eigenvalues, solid black lines denote pairs of complex eigenvalues, and red dots mark where the real part changes sign.}}
\label{viscous_eig:fig}
\end{figure}

We obtain the viscous eigenvalues using a second-order, centered finite-difference discretization of the operator $\mathcal{L}$ appropriately modified near the ends to account for the boundary conditions. The real and imaginary parts of the eigenvalue with largest real part are plotted in Figure \ref{viscous_eig:fig}(a,b). Consistent with \cite{de2017spontaneous} we find the critical strength of the follower force at which oscillations emerge in the viscous fluid is $\sigmacrit \approx 37.7$. At this critical value, the eigenvalues corresponding to the bifurcation are $\etaV\approx\pm 191 i$. We define $\omega_{0}\approx 191$ as the angular frequency of the emergent oscillation in a viscous fluid. 

Given the viscous \change{eigenvalues} as a function of $\sigma$, for a given relaxation time, $\lambda$, and  viscosity ratio, $\beta$, we identify the corresponding viscoelastic eigenvalues by solving equation \eqref{eq:ve_eig_val} for $\etaVE$. To identify instability in the viscoelastic case, we only need to compute the viscoelastic eigenvalues when the corresponding viscous eigenvalues have positive real part. We find that for $\sigmacrit <\sigma< 174.6$ there is only the single pair of eigenvalues with positive real part for the viscous fluid \change{as shown in Figure \ref{viscous_eig:fig}(c). In the remainder of this work we restrict $\sigma< 174.6$ which simplifies the stability analysis in the viscoelastic case because we only need to consider the viscoelastic eigenvalues corresponding to a single viscous eigenvalue.} 

\subsection{Asymptotic Analysis of Instability in a Viscoelastic Fluid}\label{subsec:asym}

\subsubsection{Limit of Large Relaxation Time}
As $\lambda\rightarrow\infty$ equation \eqref{eq:ve_eig_val} at leading order is 
\begin{equation}
  \etaV = \beta \etaVE + \mathcal{O}\left(\lambda^{-1}\right)
  \label{evals_asym_large_lambda:eq}
\end{equation}
Thus in this limit of large relaxation time the viscoelastic eigenvalue is proportional to the viscous eigenvalue, and two conclusions follow immediately from this relation. First, the critical follower force strength at which oscillations emerge in the viscoelastic fluid (i.e.\ the Hopf bifurcation point) is identical to the critical follower force for a viscous fluid. Second, the emergent frequencies in the two fluids are related by 
\begin{equation}
\omega_{\textrm ve}=\beta^{-1}\omega_{\textrm v}.\label{freq_asym_large_lambda:eq}\end{equation}
 Because $0<\beta\leq 1$, the frequency in the viscoelatic case is always larger than the frequency in the viscous case. In summary, in the limit $\lambda\rightarrow\infty$, the bifurcation occurs at the same follower force strength and the emergent frequency is greater in the viscoelastic case by a factor inversely proportional to the viscosity ratio. We note that many biologically relevant media such as respiratory and cervical mucus have relatively large relaxation times \cite{lai2009micro}.

\subsubsection{Limit of vanishing polymer viscosity:  $\beta\rightarrow 1$}
\label{betato1:sec}

\change{There are two limits in which the viscoelastic fluid reduces to a viscous fluid: vanishing relaxation time ($\lambda\rightarrow 0$) and vanishing polymer viscosity  ($\beta\rightarrow 1$). Both limits are useful in considering perturbation from a viscous fluid. We consider the limit $\beta\rightarrow 1$, which physically corresponds to the limit of small polymer viscosity and is typical of dilute polymeric solutions. In this limit we are able to examine how the critical follower force strength and emergent frequency depend on both the relaxation time and follower force strength for all relaxation times. As we explain later, this analysis also captures the limit $\lambda\rightarrow 0$.}

We next consider the limit $\beta\rightarrow 1$, which physically corresponds to the limit of small polymer viscosity and is typical of dilute polymeric solutions. In this limit we are able to examine how the critical follower force strength and emergent frequency depend on both the relaxation time and follower force strength.

At the bifurcation point the viscoelastic eigenvalue is pure imaginary, i.e.\ $\etaVE=i\omega$, where $\omega$ is the angular frequency at the bifurcation point. The relationship between the viscous and viscoelastic eigenvalues, equation \eqref{eq:ve_eig_val}, at the \change{bifurcation} point is
\begin{equation}
  \etaV\left(\sigma\right) = \frac{(1-\beta)i\omega} {\lambda i\omega  + 1} + \beta i\omega,
  \label{eq:ve_eig_at_bif}
\end{equation}
where \change{the notation $\etaV\left(\sigma\right)$ is used to emphasize that the viscous eigenvalue depends on the unknown follower force strength, $\sigma$.} This is a single complex valued equation involving the two real-valued unknowns $\sigma$ and $\omega$. 

For $\beta$ close to 1, $\sigma$ will be close to $\sigma_{0}$, the critical follower force strength in a viscous fluid. We linearize $\etaV$ about this point so that 
\begin{equation} 
	\etaV\left(\sigma\right)=i\omega_{0} + \hat{\sigma}(a+bi) + \mathcal{O}\left(\hat{\sigma}^2\right),
	\label{eq:linearization_etaV}
\end{equation}
where $\hat{\sigma}=\sigma-\sigma_{0}$ and $\left.(d\etaV/d\sigma)\right|_{\sigma_{0}}=a+bi$. 
Eliminating $\etaV$ from \eqref{eq:ve_eig_at_bif} using \eqref{eq:linearization_etaV} and then equating the real and imaginary parts results in
\begin{align}
   a\hat{\sigma} & = \frac{\epsilon \omega^{2}\lambda}{1 + \omega^{2} \lambda^{2}} 
                     + \mathcal{O}(\hat{\sigma}^2),
    \label{sigma_asym_near_bif:eq}                 \\
   \omega_{0} + \hat{\sigma}b &= 
        \omega \bigg(1 - \epsilon \frac{\omega^{2}\lambda^{2}}{1 + \omega^{2}\lambda^{2}}\bigg)
        + \mathcal{O}(\hat{\sigma}^2),
    \label{omega_asym_near_bif:eq} 
\end{align}
where $\epsilon=1-\beta$. We seek a solution in the limit $\epsilon \rightarrow 0$ by using the expansions $\hat{\sigma} = \epsilon \sigma_{1} + \epsilon^{2} \sigma_{2}\ldots$ and $\omega = \omega_{0} + \epsilon \omega_{1} + \ldots$ and matching the $\mathcal{O}(\epsilon)$ terms. The resulting expansions for critical follower force and corresponding frequency are
\begin{align}
\sigma & = \sigma_{0} + (1-\beta)\frac{ \lambda \omega_{0}^2 }{a\left(1 + \lambda^{2}\omega_{0}^2\right)} + \mathcal{O}\bigl((1-\beta)^2\bigr), 
\label{sigma_asym_small_beta:eq} \\
\omega & = \omega_{0} + (1-\beta)\omega_{0}\left(\frac{\frac{b}{a}\lambda \omega_{0} + \lambda^{2}\omega_{0}^2}{1 + \lambda^{2}\omega_{0}^2}\right) + \mathcal{O}\bigl((1-\beta)^2\bigr).
\label{omega_asym_small_beta:eq}
\end{align}
In these expressions, the relaxation time appears paired with the frequency in the product $\lambda\omega_{0}$. This quantity is similar to the Deborah number, but the frequency is fixed at the emergent frequency in the viscous limit. In this section we consider how the bifurcation location and emergent frequency depend on the scaled relaxation time $\lambda\omega_{0}$.

\change{Before continuing, we remark that one can recover the $\lambda\rightarrow 0$ expansions from these expressions. Equations  \eqref{sigma_asym_near_bif:eq} and \eqref{omega_asym_near_bif:eq} are valid near the viscous bifurcation point, and thus when $\lambda\rightarrow 0$ for all $\beta$. Equation \eqref{sigma_asym_small_beta:eq} follows directly from \eqref{sigma_asym_near_bif:eq}. The expressions for the frequency in the limit $\lambda\rightarrow 0$ that one obtains from expanding \eqref{omega_asym_near_bif:eq} or \eqref{omega_asym_small_beta:eq} are equivalent at $\mathcal{O}(\lambda)$.}

\begin{figure}[b]
\centering
\includegraphics[width=0.375\textwidth]{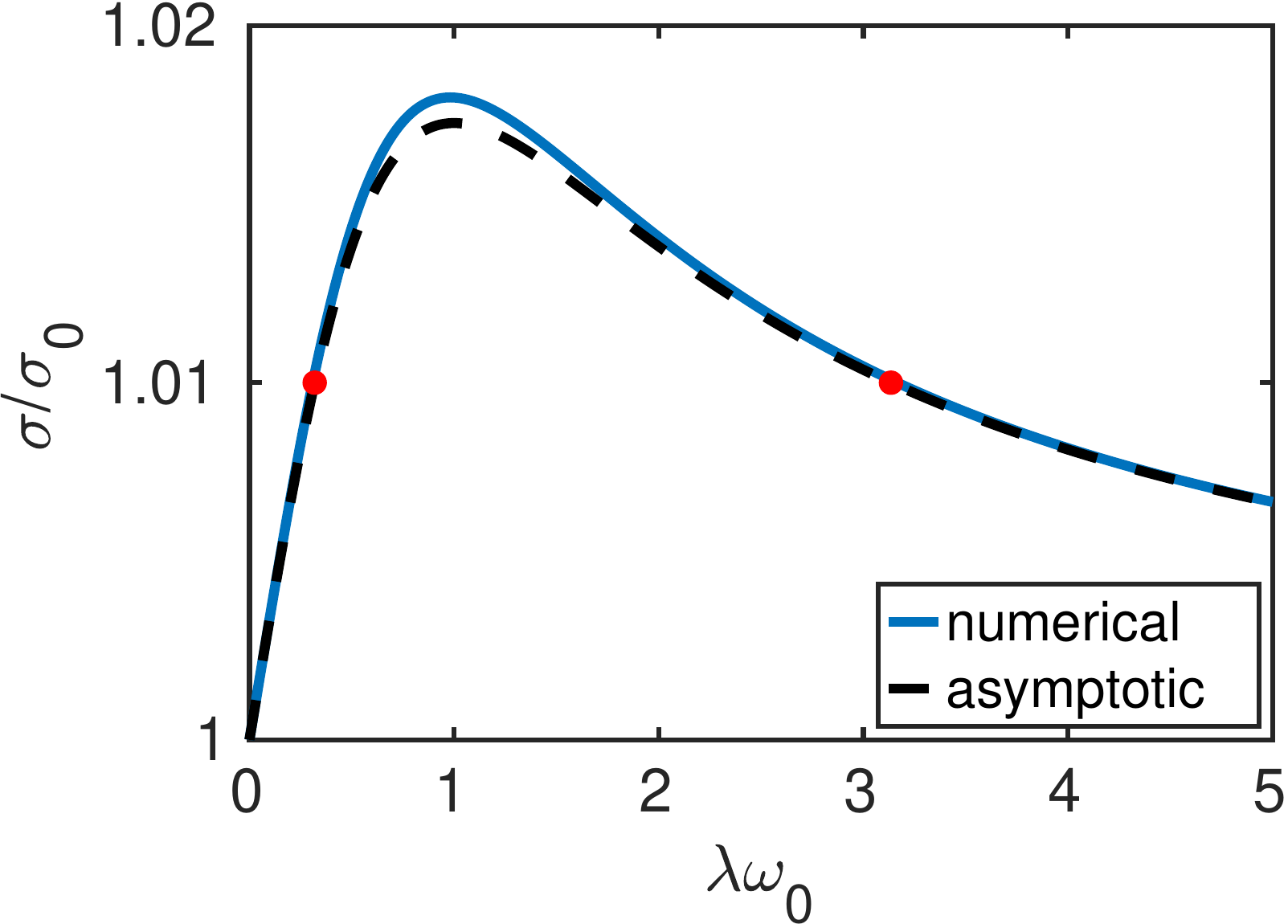}
\includegraphics[width=0.375\textwidth]{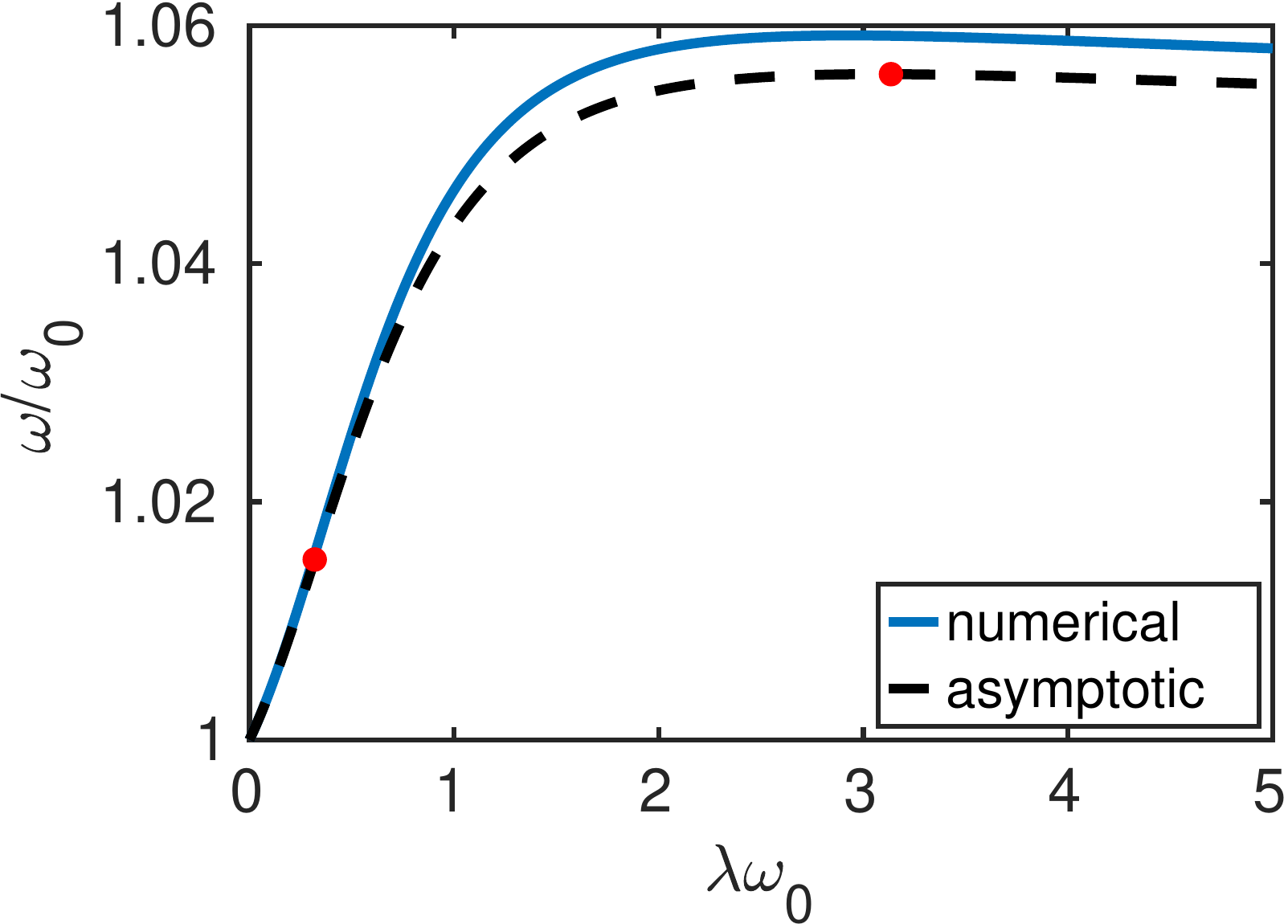}
\caption{Plots of the asymptotic \change{and numerical} solutions as $\beta\rightarrow 1$ for the (a) critical follower force and (b) emergent frequency at the bifurcation as functions of the relaxation time for $\beta=0.95$. The two red dots mark the bifurcation points and emergent frequency, respectively, for $\sigma=1.01\sigma_{0}$.}
\label{asym_sigma_freq:fig}
\end{figure}

In Figure \ref{asym_sigma_freq:fig} we plot the asymptotic \change{and numerical} solutions for critical follower force and emergent frequency in a viscoelastic fluid relative to their respective values in a viscous fluid for $\beta=0.95$ as a function of relaxation time. \change{Plots for other values of $\beta$ are shown in the Supplementary Information.} As the relaxation time increases, the critical force increases and then decreases. The peak critical force occurs at  $\lambda\omega_{0}=1$, and as expected from the large relaxation time analysis $\sigma\rightarrow\sigma_{0}$ as $\lambda\rightarrow\infty$. 

The emergent frequency also increases and then decreases as it approaches to a frequency greater than the corresponding viscous frequency, again, as expected. However, note that  $\omega/\omega_{0}\rightarrow \beta^{-1}$ according to \eqref{evals_asym_large_lambda:eq} and $\omega/\omega_{0}\rightarrow 1+(1-\beta)$ according to \eqref{omega_asym_small_beta:eq}. Because $\beta^{-1}=1+(1-\beta)+\mathcal{O}(1-\beta)$, these expressions are not inconsistent. The peak in the frequency occurs at a larger relaxation time than the peak in the critical force. Specifically the peak occurs at $\lambda\omega_{0}=a/b +\sqrt{a^2/b^2+1}\approx 3.07$.

For small polymer viscosity, there remains a single critical value of the follower force above which oscillations emerge. However, for the range of follower forces below the peak in Fig.\ \ref{asym_sigma_freq:fig}(a) (which is approximately $1<\sigma/\sigma_{0}<1+(1-\beta)/(2a\sigma_{0})$), there are two Hopf bifurcation points as the relaxation time changes. For a follower force in this range, the filament oscillates at low and high relaxation times while the rest state is stable for an intermediate range of relaxation times.  The frequency at the higher relaxation time is generally greater than that of the lower relaxation time. For example, for $\beta=0.95$ and $\sigma=1.01\sigma_{0}$ the two bifurcation points and critical frequencies are marked with red dots in Figure \ref{asym_sigma_freq:fig}. In later sections, we examine how the frequency changes with relaxation time for fixed follower force, and we will see that generally the frequency increases with increasing relaxation time.   


\subsection{Bifurcation Location and Emergent Frequency for General $\beta$}


\begin{figure}[b]
\centering
\includegraphics[width=\singlepanelwidth]{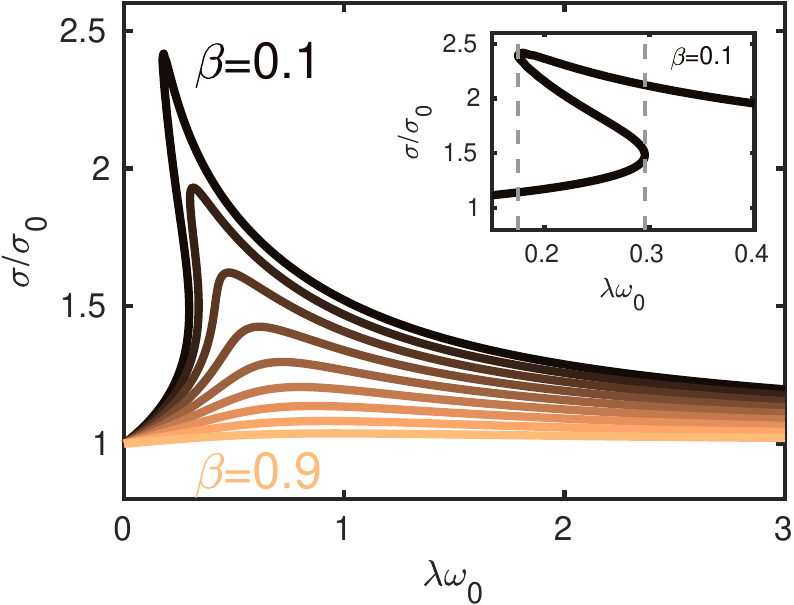}
\caption{Location of the bifurcation point in the $\sigma$\--$\lambda\omega_{0}$ plane for $\beta=0.1, 0.2,\ldots 0.9$. Inset: Location of the bifurcation for $\beta=0.1$ zoomed in to show that between the two dashed vertical lines there are three bifurcation points as the follower force changes.
}
\label{bif_loc:fig}
\end{figure}

We examine the bifurcation for general values of the viscosity ratio $\beta$ by solving equation \eqref{eq:ve_eig_val} for $\etaVE$ using the numerically computed viscous eigenvalues. Note that there are two values of $\etaVE$ for each viscous eigenvalue. We found that for all parameter regimes we explored, one of the two values of $\etaVE$ always had negative real part. \change{As discussed in the Supplementary Information, this additional eigenvalue scales with $1/\lambda$ and is likely related to the fluid relaxation time scale.} In Figure \ref{bif_loc:fig} we show the location of the bifurcation in the $\sigma$\--$\lambda\omega_{0}$ plane for $\beta=0.1, 0.2,\ldots 0.9$. For many values of $\beta$ the shape of the curve denoting the location of the bifurcation is qualitatively similar to that of the asymptotic result. In the limit $\beta\rightarrow 1$, the peak in the critical follower force  occurs at $\lambda\omega_{0}=1$, but as $\beta$ decreases the relaxation time corresponding to this peak decreases. 

For example for $\beta=0.5$, the peak occurs near $\lambda\omega_{0}\approx 0.72$. For small values of $\beta$ the critical force strength is no longer a single valued function of the relaxation time. The inset in Figure \ref{bif_loc:fig} shows the bifurcation location for $\beta=0.1$ for a smaller range of $\lambda$ to highlight this feature. At $\beta=0.1$ for $0.1749<\lambda\omega_{0}<0.2955$ (end points marked with gray lines in the figure), there are three $\sigma$ values at which bifurcations occur.

\begin{figure}[t]
\centering
\includegraphics[width=\singlepanelwidthwide]{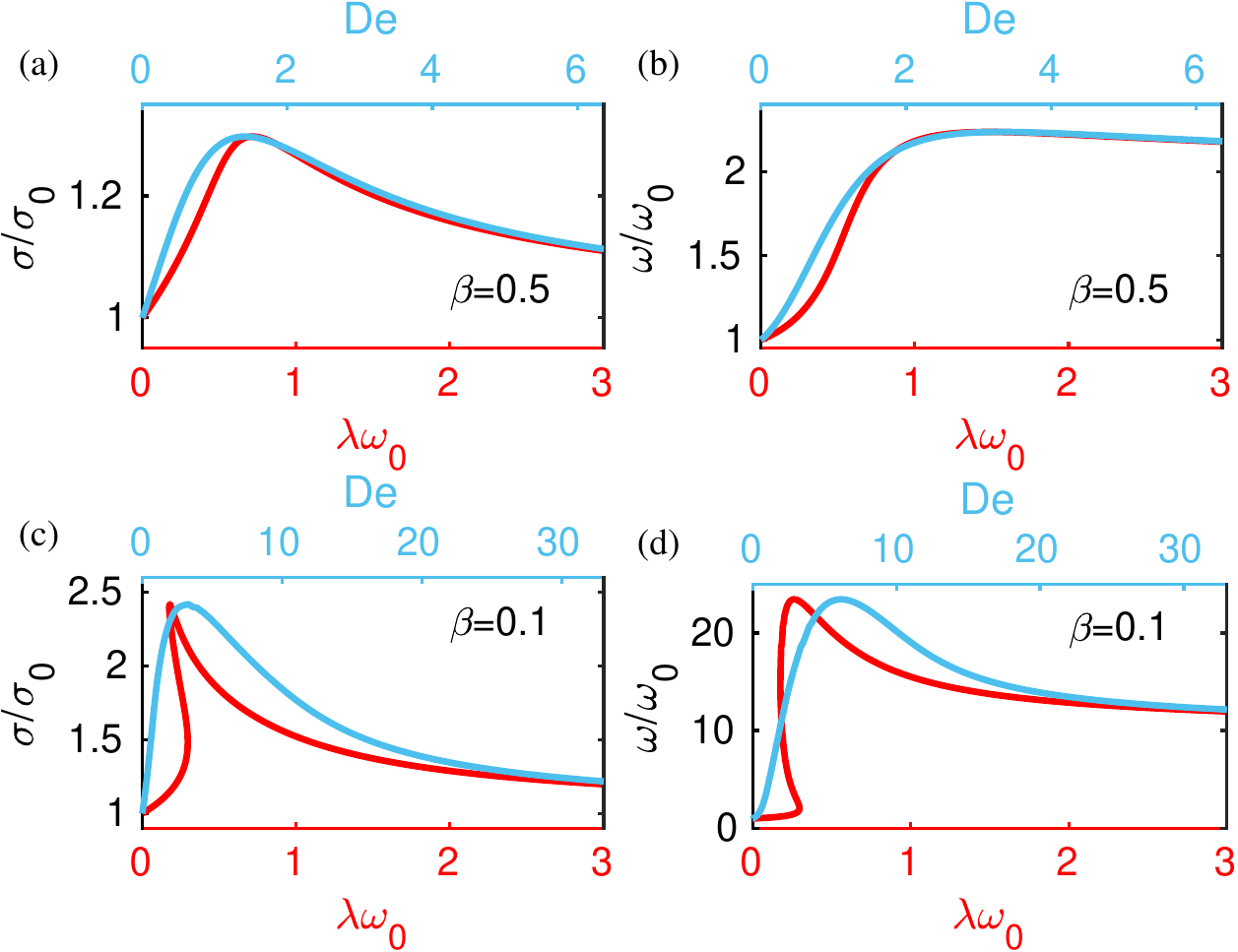}
\caption{Bifurcation location (a,c) and frequency (b,d) at the bifurcation for $\beta=0.5$ (a,b) and (c,d) $\beta=0.1$. The two curves in each panel represent two different scalings for the relaxation time, $\lambda$. The red curves (bottom axes) show how the data depend on $\lambda\omega_{0}$, where $\omega_{0}$ is the angular frequency at the bifurcation point in a viscous fluid. The blue curves (top axes) show the data depend on the Deborah number $\text{De}=\lambda\omega$, where $\omega$ is the emergent frequency in a viscoelastic fluid. Because the emergent frequency depends on the relaxation time, De represents a nonuniform scaling of the relaxation time.}
\label{bif_freq_de_and_de0:fig}
\end{figure}

The quantity $\lambda\omega_{0}$ that appears in the asymptotic expressions \eqref{sigma_asym_small_beta:eq}\--\eqref{omega_asym_small_beta:eq} represents the Deborah number in the limit of vanishing polymer viscosity ($\beta\rightarrow 1$). More generally, we take as the Deborah number the product of the relaxation time and the emergent frequency; i.e.  $\DE=\lambda\omega$. In Figure \ref{bif_freq_de_and_de0:fig} we examine how the critical force and the emergent frequency depend on $\lambda\omega_{0}$ and $\DE=\lambda\omega$ for $\beta=0.5$ and $\beta=0.1$. For $\beta=0.5$ the shapes of the critical force and emergent frequency are qualitatively similar when viewed as functions of either $\lambda\omega_{0}$ or $\DE$. However for $\beta=0.1$, there is a substantial difference in these curves. The critical force and emergent frequency are single-valued functions of $\DE$. Also, the shapes of these curves as functions of $\DE$ are qualitatively similar to the shapes predicted by the asymptotic analysis as $\beta\rightarrow 1$ and corresponding curves for $\beta=0.5$. 
\change{Both the critical force and the emergent frequency are multivalued in the same range of relaxation times ( $0.1749<\lambda\omega_{0}<0.2955$ for $\beta=0.1$). For each value of $\lambda$ in this range, there are three different bifurcation points each with their own frequency, and hence their own distinct Deborah number. This explains why the critical force and emergent frequency are single-valued functions of $\DE$.}

In later results we use both the scaled relaxation time $\lambda\omega_{0}$ and the Deborah number $\lambda\omega$. The former is particularly useful when examining how quantities change with relaxation time. Because the emergent frequency depends on the relaxation time the Deborah number is not simply proportional to the relaxation time.

%
%
\section{Frequency Analysis for a Fixed Follower Force in a Viscoelastic Fluid}


\subsection{Frequency changes for varying relaxation times}

\begin{figure}
\centering
\includegraphics[width=\singlepanelwidth]{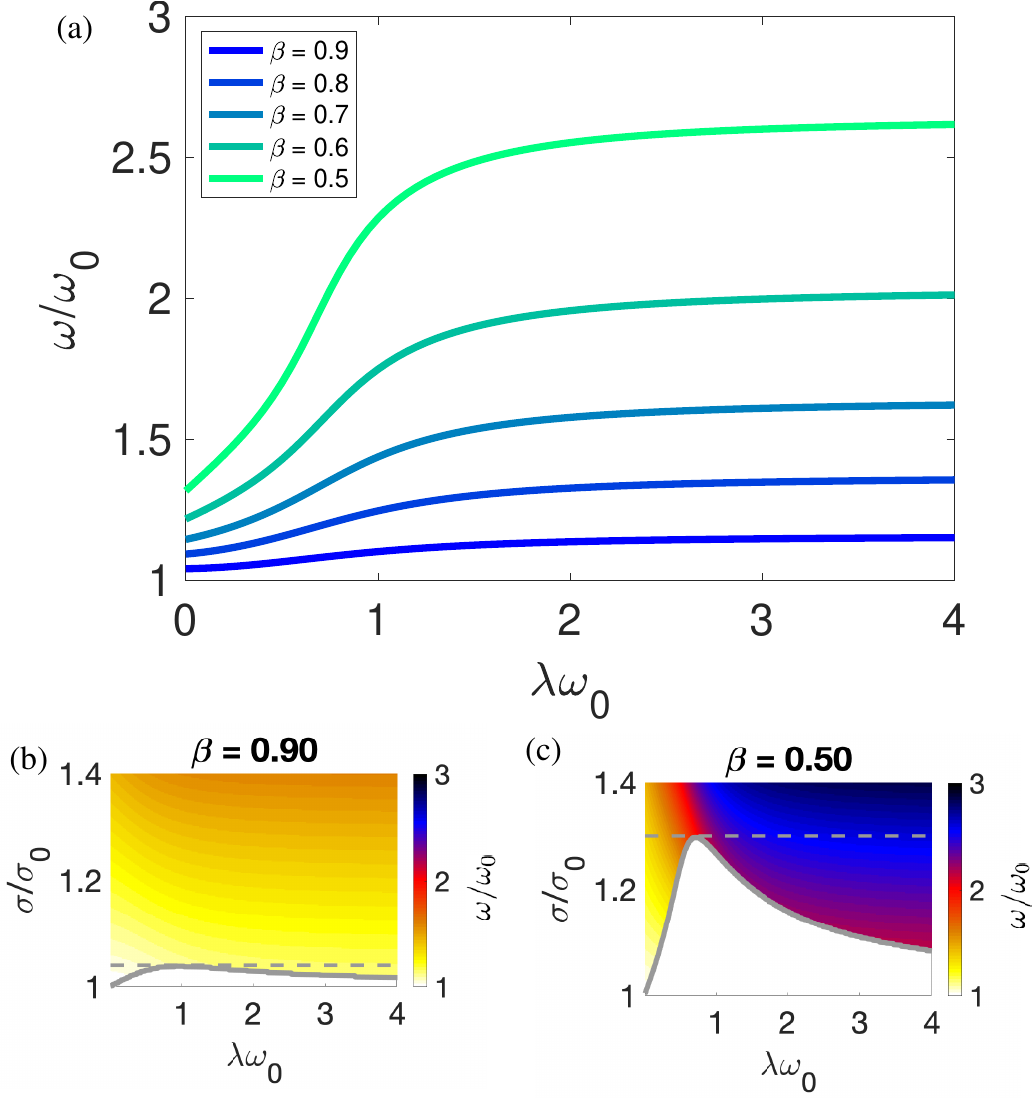}
\caption{Emergent frequency (a) as a function of scaled relaxation time for $\beta\ge 0.5$ at a fixed follower force that is 1\% higher than the maximum force at the bifurcation for the corresponding $\beta$.  Color fields of frequency for $\beta=0.9$ (b) and $\beta=0.5$ (c). Solid lines show location of bifurcation, and dashed line shows fixed force value for corresponding frequency values shown above. }\label{fig_vbeta}
\end{figure}

In the previous section, we examined how the emergent frequency at the bifurcation depends on the fluid parameters. However, the bifurcation location depends on the fluid parameters. Hence as the fluid parameters vary the follower force varies as well. Here we explore how the emergent frequency depends on fluid elasticity and viscosity at a \textit{fixed} follower force. Close to the bifurcation the angular frequency is approximately the imaginary part of the eigenvalue with positive real part. In order for the results from linear stability analysis to be relevant, we consider solutions that are close to the bifurcation. For a fixed  $\beta$, the critical force where oscillations emerge is non-monotonic in $\lambda,$ and there is a maximum critical force as a function of $\lambda$; e.g.\ see  Fig.\ \ref{bif_loc:fig}. We choose the force to be approximately $1\%$ higher than the maximum for a particular value of $\beta$ and examine how the frequency changes as a function of $\lambda$. This choice of force is large enough to avoid bifurcations in $\lambda$ where the oscillations cease and small enough to avoid large amplitude motion where the linear analysis is less accurate. For example, for $\beta\ge 0.5$ the local maximum in force $\sigma\lesssim 1.3\sigma_0$ which is still relatively close to the bifurcation for all relaxation times. 
  
In Fig. \ref{fig_vbeta} (a) we show the emergent frequency scaled by $\omega_0$  at a fixed follower force as a function of the scaled relaxation time $\lambda\omega_0$ for a range of $\beta\ge 0.5.$ The emergent frequency is monotonically increasing for a fixed follower force for each $\beta$ and the frequency increase levels off for large $\lambda\omega_0$. The emergent frequency also increases with decreasing $\beta.$ For $\beta=0.9$ the frequency at $\lambda\omega_0=4$ is about 10\% higher than the viscous frequency at the same force, whereas for $\beta=0.5$ the frequency at $\lambda\omega_0=4$ is nearly double the viscous frequency at the same force. 

In Fig. \ref{fig_vbeta} (b) and (c) we  show color-fields of the emergent frequency as a function of both $\lambda\omega_0$ as well as $\sigma/\sigma_0$ for $\beta=0.9,0.5.$ The fixed follower force strength corresponding to the figure above are highlighted with grey dashed lines. For  follower forces fixed at higher values (above the grey line), the qualitative behavior of the frequency is the same, namely the frequency increases rapidly for $\lambda\omega_0\lesssim 1$ and levels off for higher relaxation times.  Quantitatively, higher forces  lead to higher frequencies overall. 


\subsection{Comparing analysis and numerical simulations}
\begin{figure}[h]
\centering
\includegraphics[width=0.65\textwidth]{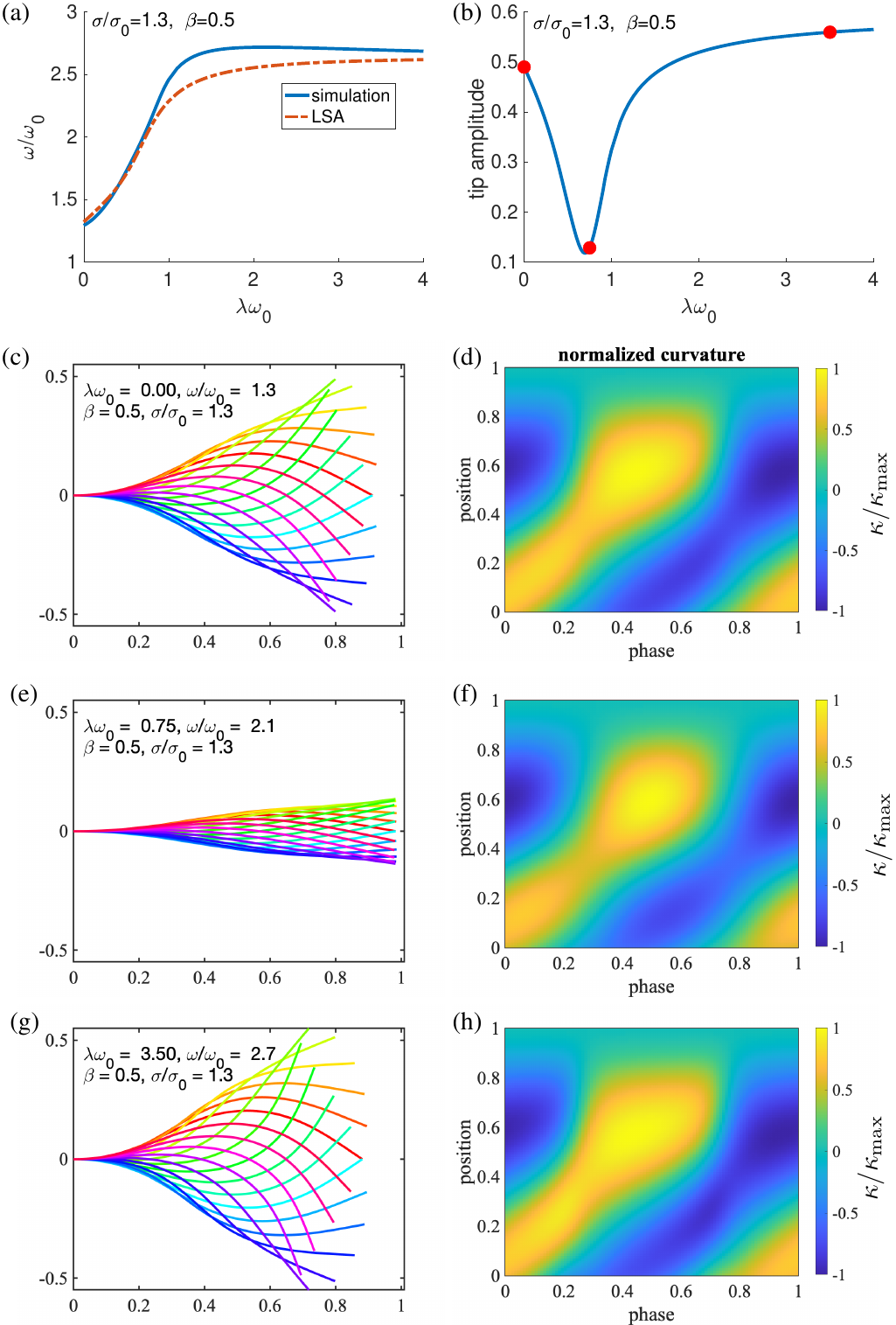}
\caption{Emergent frequency (a) and tip amplitude (b) for $\beta=0.5,$ $\sigma/\sigma_0=1.3.$ Frequency figure shows a comparison with simulation and linear stability analysis. Emergent shapes \change{with 20 snapshots per period} (c,e,g) and curvature normalized by its maximum \change{($\kappa/\kappa_{\textrm{max}}$) }(d,f,h)  over a period for simulations with $\lambda\omega_0=0,0.75,3.5$, these values are highlighted with red dots above. \change{The maximum values of curvature for these three values of $\lambda\omega_0$ are  $3.01$, $1.15$, and $3.56$, respectively. The amplitude of the curvature follows the same trend as the tip amplitude.} 
}
\label{fig_shapes}
\end{figure}

To explore how well the linear stability analysis predicts the emergent frequencies away from the bifurcation we solve Eqs.\ \eqref{eq:model1} \-- \eqref{eq:model2} numerically with boundary conditions given by Eqs.\ \eqref{eq:bc2} \-- \eqref{eq:bc3}, which accounts for both normal and tangential deformations. Details of the numerical method are described in the Supplementary Information. 
%
With these simulations we are also able to examine how the amplitude and shape change with varying fluid rheology.

%
In Fig.\ref{fig_shapes} (a) we compare the results for emergent frequency in the linear stability analysis and the simulations for $\beta=0.5,\sigma/\sigma_0=1.3.$  
\change{The corresponding amplitude of the tip of the filament is shown in Fig.\ref{fig_shapes} (b). Despite the fact that the amplitude is not particularly small, the simulations qualitatively match the frequency predicted from the linear stability analysis, with the simulations exhibiting slightly higher frequencies for moderate $\lambda\omega_0.$ 
The amplitude of the oscillation is not available from the linear analysis, but the amplitude is related to the distance from the bifurcation. Because the oscillation emerges at a Hopf bifurcation, the amplitude should grow like $(\sigma-\sigma_{c})^{1/2}$, where $\sigma_{c}$ represents the follower force strength at the bifurcation. 
The fact that the amplitude initially decreases and then increases to a constant value as $\lambda$ increases is consistent with how the distance to the bifurcation changes.}

In Fig.\ref{fig_shapes} (c-h)  shapes (left) and normalized curvatures (right) for the simulations at $\lambda\omega_0=0,0.75, 3.5$ (corresponding to the red dots on the amplitude figure above) are shown. The plots in Fig.\ref{fig_shapes} (c,e,g) show the filaments at the same phase in the period (phase is labeled by color) and the actual frequency of motion is listed in the figure. Other than changes to amplitude and frequency the overall shapes are similar. This can be seen more clearly  in the kymographs of curvature normalized by its maximum value in Fig.\ref{fig_shapes} (d,f,g). \change{The peak values of curvature are  $3.01$, $1.15$, and $3.56$ for $\lambda\omega_0=0$, $0.75,$ and $3.50$, respectively. The amplitude of the curvature follows the same trend as the tip amplitude, which as noted above is related to the distance from the bifurcation.} The curvatures exhibit subtle differences but are similar.  

Although one may expect more significant shape changes as a function of rheology, the shape of the oscillating filament near the bifurcation is determined by the eigenfuctions of the operator $\mathcal{L}$ in \eqref{vis_eig_prob:eq}. These eigenfunctions do not depend on the fluid elasticity ($\lambda$ nor $\beta$). It is only the eigenvalues of the system that change with fluid elasticity.
We only considered parameters near the bifurcation point to remain in the low amplitude regime so that linear viscoelasticity was a reasonable approximation. Other analyses of filaments subject \change{to} follower forces in viscous fluids demonstrated significant shape variation and different kinds of motion at large amplitudes that depend  on the boundary conditions and distribution of follower forces \cite{Feng:2018:RSI:instabilityoscillations,Fily:2020:RSI:bucklinginstabilities}. 
There may be shape changes due to fluid elasticity at higher amplitudes, but at large amplitude one must consider viscoelastic nonlinearity \cite{thomases2017role,thomases2019polymer}.

\subsection{Comparing \change{frequency changes} with experiments}

In \cite{qin2015flagellar} the flagellar beat pattern, beat frequency, and swimming speed of the  biflagellated alga \textit{C.\ reinhardtii} were measured in response to systematic variation of the fluid viscosity and relaxation time (or fluid elasticity). Surprisingly, it was observed that the beat frequency increased with increasing relaxation time. In a Newtonian fluid the beat frequency decreased monotonically with the fluid viscosity, but in viscoelastic fluid the frequency changed nonmonotoically with the fluid viscosity \--- initially decreasing, then increasing and appearing to plateau. The physical origins of the frequency response to fluid elasticity are not currently understood. Here we examine the predictions of the follower model to compare with these experimental observations.

%
\change{
It is reasonable to ask whether one expects the predictions from the analysis of the follower model to be relevant to flagella which are driven by dynamic motor forces along the filament. 
If the motor activity is constant or does not change with the rheology of the fluid, then the predictions of our analysis that follow from the relationship between the eigenvlaues in eq.\ \eqref{eq:ve_eig_val} will hold. Namely, the analysis in \S\ref{Sec:LSA} predicts a higher beat frequency in a viscoelatic fluid which approaches a constant as the relaxation time increases.
Several studies have examined different mechanisms of motor feedback and control in the linearized equations, and these models are capable of matching experimental data \cite{Camalet_2000,Riedel-Kruse:2007:HSFP:motorsshape,Sartori:2016:Elife:dynamiccurvature}.  
While there are different proposed control mechanisms, all of them are assumed to depend on the frequency of the emergent beat.
In the Supplementary Information we show that even when the motor activity changes with emergent frequency, in the limit of vanishing polymer viscosity ($\beta\rightarrow 1$), the expression for the frequency at the bifurcation is of the same form as \eqref{omega_asym_small_beta:eq}. Thus it is expected that the frequency generally increases with relaxation time and approaches a constant value in the limit of large relaxation time. 
As we show below, these two features of the frequency response to fluid elasticity are consistent with the data from \cite{qin2015flagellar}, and thus our analysis may provide insight into the mechanisms underlying the observations. 
}


%
%
%
%
%
%
%
%
%
%
%

%
The viscoelastic fluids in \cite{qin2015flagellar} were prepared by adding small amounts of the high molecular weight, flexible polymer polyacrylamide to water. The addition of polymer increases the fluid relaxation time, but it also changes the total viscosity. In terms of the parameters used in this work, both $\beta$ and $\lambda$ change simultaneously.    

In order to compare the predictions of the follower model with these experiments, we fit the rheological data from \cite{qin2015flagellar} to find functional forms for the polymer viscosity, $\mu_{p}(c)$, and the (dimensional) relaxation time, $\tau(c)$, as functions of the polymer concentration $\c$ in ppm. Details of our fitting procedure are given in the Supplementary Information.
 Using these fits and the non-dimensionalization in Sec.\ \ref{Sec:Model}, we obtain the dimensionless relaxation time and solvent fraction $\lambda(c)=\tau(c) k_b/(L^4(\mu_s+\mu_p(c)))$ and  $\beta(c)=\mu_s/(\mu_s+\mu_p(c))$, respectively, as functions of the polymer concentration. 

\begin{figure}[t]
\centering
\includegraphics[width=\singlepanelwidthwide]{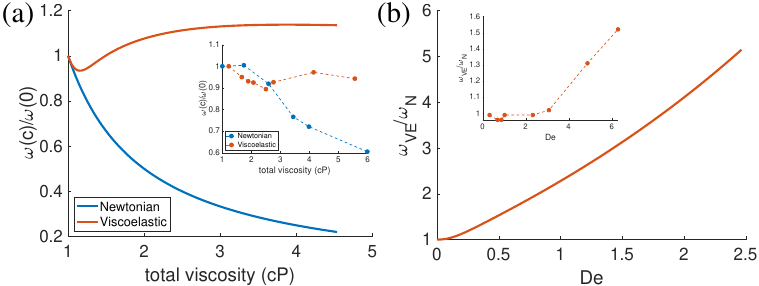}
\caption{Frequency scaled by its value at $c=0$ as a function of the total viscosity (a). Newtonian viscosity follows 1/viscosity scaling for comparison. Frequency relative to viscous frequency as a function of the Deborah number (b). Simulations run using fixed value of force $\sigma/\sigma_0=1.5,$ and $k_b /L^4=0.25.$ Inset graphs using data from \cite{qin2015flagellar}}\label{fig_vcc}
\end{figure}

We use these models for $\lambda(\c)$ and $\beta(\c)$ in simulations for a fixed follower force $\sigma/\sigma_0=1.5,$ and $k_b/L^4=0.25$.  Other choices for $\sigma/\sigma_0$ and $k_b/L^4$ are considered in the Supplementary Information.
We vary the concentration over the range $\c=0-80$ ppm and compute the dimensional frequency of the oscillation, $\omega_{VE}(c)$. In Fig.\ \ref{fig_vcc}(a) we plot this frequency normalized by the frequency at $\c=0$ as a function of the total viscosity, $(\mu_{s}+\mu_{p})(c).$ We compare these results with the experimental data from \cite{qin2015flagellar}  plotted in the inset. It is remarkable that frequency changes in a viscoelastic fluid predicted by the follower model agree qualitatively with the experimental data. Specifically, the non-monotonicity of the frequency dependence on viscosity as well as the plateau for high viscosity are all captured by the follower model.

In a Newtonian fluid, the frequency of oscillations in the follower model at fixed follower force is inversely proportional to the viscosity. This is because in the dimensionless equations the strength of the follower force is the only parameter, and the time scale in the nondimensionalization is proportional to the viscosity. On Fig.\ \ref{fig_vcc}(a) we include a plot of the normalized oscillation frequency in a Newtonian fluid: $\omega_{N}(c)/\omega_{N}(0)=(\mu_s+\mu_{p})^{-1}$, and the corresponding experimental data is shown in the inset. Both the follower model and the experimental data show that the frequency decreases with increasing viscosity. However, the follower frequency is inversely proportional to viscosity, but as discussed in  \cite{qin2015flagellar}, the measured frequency in Newtonian fluid scaled like $(\mu_s+\mu_{p})^{-1/2}$ for large viscosity, which is consistent with the frequency scaling predicted by a model that includes force-sensitive dynein motor activity \cite{Camalet_2000}. 

The non-monotonic response of the frequency to viscosity can be explained using the asymptotic analysis in Sec.\ref{subsec:asym}(a). For low polymer concentrations, $\beta$ is close to 1, and the asymptotic expression for the frequency in Eq.\ \eqref{omega_asym_small_beta:eq} holds. Redimensionalizing this expression, the frequency decrease from increasing viscosity occurs at first order in concentration, but the frequency increase from elasticity occurs at second order in concentration. Therefore for small concentrations, the frequency should drop at the same rate as the frequnecy in a viscous fluid, as is observed in Fig.\ \ref{fig_vcc}(a). Further, at high polymer concentration the relaxation time is large, and the frequency is approximately given by Eq.\ \eqref{freq_asym_large_lambda:eq}. Because $\beta$ grows at the same rate that the time scale decays, when Eq.\ \eqref{freq_asym_large_lambda:eq} is redimensionalized it predicts that $\omega_{VE}(c)\sim\omega_{VE}(0)$; i.e.\ the frequency should approach the frequency at zero polymer concentration. In Fig.\ \ref{fig_vcc}(a) we see the frequency appears to approach a value 10\% higher than predicted, but recall that Eq.\ \eqref{freq_asym_large_lambda:eq} holds at the bifurcation point, and the results in Fig.\ \ref{fig_vcc} are away from the bifurcation. In summary, the analysis predicts that the frequency should initially drop at low polymer concentrations when viscous effects dominate, but it must eventually increase and approach a constant in a regime in which viscous and elastic effects counter balance each other. 
These same trends were observed in experiments \cite{qin2015flagellar}, and thus this analysis provides a \change{possible} mechanistic understanding of the observed frequency response to changes in viscosity in viscoelastic fluids.  


Viscosity can be varied for both Newtonian and viscoelastic fluids. In order to isolate the effects of elasticity on frequency in \cite{qin2015flagellar} 
the frequency was measured for both Newtonian and viscoelastic fluids at the same total viscosity. The  frequency in a viscoelastic fluid relative to the  frequency of a viscous fluid of the same viscosity was reported based on Deborah number $\DE = \lambda\omega.$ Similarly, in Fig.\ \ref{fig_vcc}(b) we plot the frequency of the oscillation in a viscoelastic fluid normalized by the corresponding frequency in a Newtonian fluid of the same viscosity as a function of $\DE.$  We contrast these results with the results in Fig.\ \ref{fig_vbeta} for the frequency as a function of the relaxation time for fixed $\beta$. In both cases, the frequency increases as the relaxation time increases, but when the viscosity is fixed, as in  Fig.\ \ref{fig_vbeta}, the frequency levels off for high relaxation time. When the relaxation time and viscosity change together (via the polymer concentration), as in Fig.\ \ref{fig_vcc}(b), the frequency does not approach a constant.  These results are, again, qualitatively  similar to the experiments from \cite{qin2015flagellar} (see inset).  
%
%
The agreement between the follower model and the data on \textit{C.\ reinhardtii} is remarkable given that flagella are powered by dynamic internal molecular motors and the follower model we analyzed is driven by a fixed-strength external force.  This suggests that the frequency response to fluid elasticity that arises from changes in the fluid drag may apply more generally to systems with different driving forces.

%
%
\section{Discussion}

%
Despite the fact that fluid rheology is known to affect the shape and frequency of beating flagella in many biological systems, there is limited data in which the fluid elasticity is systematically varied, and a mechanistic understanding of how fluid elasticity affects emergent motion does not exist. Theoretical explorations have shown how fluid elasticity can change the shape of the beat with prescribed active forces \cite{fu2008beating,thomases2017role}, but these models cannot be used to understand the frequency changes observed in experiments \cite{qin2015flagellar}. In this paper we extend the model of an elastic filament driven by a follower force at the tip from \cite{de2017spontaneous} to examine how fluid elasticity affects the emergence of oscillations and their resulting frequency.

%
As in a viscous fluid \cite{de2017spontaneous}, there is a Hopf bifurcation at a critical force at which beating emerges. Our analysis identified how the bifurcation location depends on the relaxation time and viscosity ratio. In a viscoelastic fluid the force required to induce oscillations is always higher than in a viscous fluid.   Moreover,  there are parameter regions where increasing fluid relaxation time  will stabilize  an  oscillating filament, but upon  further increase in the relaxation time the filament will again oscillate at a higher frequency. Our analysis predicts that fluid elasticity generically increases the frequency of beating over the same filament in a viscous fluid at the same force in agreement with the experimental observations in \cite{qin2015flagellar}.  When the relaxation time and total viscosity increase in tandem through the polymer concentration, competing effects of elasticity and viscosity lead to a non-monotonic response of frequency on viscosity that again agrees well with experiments \cite{qin2015flagellar}.

%
\change{
In \cite{qin2015flagellar} it was observed that although the frequency of the beat increased in viscoelastic fluids the swimming speed decreased. The shape of the beat changed significantly in viscoelastic fluids, namely the maximum of the flagellum curvature increased, but the bending at the basal end was reduced. In \cite{li2017flagellar} we performed numerical simulations of swimmers based on the gaits from  \cite{qin2015flagellar} to separate the effects of changes in gait and changes in fluid rheology on the swimming speed.  This work showed that the reduction in speed resulted from both the change of the shape of the beat and the nonlinear growth of elastic stress around the flagella. 
The model analyzed in this work does not predict shape changes in response to viscoelasticity. At low amplitude the shape is determined by the eignefunctions of the linearized operator, which do not depend on the fluid elasticity. Although the frequency response predicted from our analysis is consistent with \cite{qin2015flagellar}, the reported shape changes in the flagella beat in response to fluid elasticity cannot be captured with the model analyzed here.
Capturing shape changes due to fluid elasticity requires a more sophisticated model of the active forces from molecular motors.
}

%
%
The model we analyzed did not include mechanical feedback on the driving force. There are many theories about how mechanical feedback on molecular motors leads to spatiotemporal coordination of motor activity to produce the flagellum beat, but it has also been shown that the motor coordination is not necessary for producing the beat \cite{bayly2016steady}. Even if the mechanical feedback on motor activity is not responsible for coordination, motor regulation could play a role in modulating the flagellum beat. \change{Our analysis captures} the  qualitative changes in the frequency observed in \cite{qin2015flagellar} in response to fluid elasticity, but quantitative agreement between the model and data would likely require a more sophisticated model that includes dynamic motor activity. For example, our analysis predicts that in a Newtonian fluid, the frequency is inversely proportional to the viscosity (see also \cite{bayly2016steady}), but as discussed in \cite{qin2015flagellar}, the motor model from \cite{Camalet_2000} predicts that the frequency is inversely proportional to the square root of the viscosity, which was a better fit to the data (see Fig. \ref{fig_vcc}). 

%
One approach to analyzing how the shape and frequency of the flagellum beat is shaped by motor models that incorporate mechanical feedback is to examine time-periodic solutions of linearized equations \cite{Camalet_2000,Riedel-Kruse:2007:HSFP:motorsshape,Sartori:2016:Elife:dynamiccurvature,bayly2015analysis}. This approach is equivalent to examining the solutions at the bifurcation point as we have done here. The form of linearized equations, and thus the eigenvalues of the corresponding operator, depends on the motor model, but the eigenvalues in a viscous fluid are related to those in a viscoelastic fluid by \eqref{eq:ve_eig_val}. This suggests that the effect of fluid elasticity on frequency discussed here may be a generic effect in  models of flagella which  incorporate feedback or regulation from molecular motors.

\begin{acknowledgments}
This work was supported in part by NSF Grant No.\ DMS-1664679 to RDG and BT, and NSF DMS Postdoctoral Fellowship award 2103380 to KGL.

\end{acknowledgments}

%
%

\clearpage
\appendix
\renewcommand{\appendixname}{Supplementary Info}

%
\renewcommand{\thefigure}{S\arabic{figure}}
\renewcommand{\theequation}{S\arabic{equation}} 
\renewcommand{\thesection}{\Alph{section}}

\setcounter{figure}{0}
\setcounter{equation}{0}
\setcounter{section}{0}


\section{Numerical Methods}
\label{app:numerics}

We express the nondimensional model equations \eqref{eq:model1}\--\eqref{eq:model2} in the form 
\begin{align}
&\beta \C {\X}_{t} =  \F({\bf{X}}) + \Fp, \\
&\lambda \Fp_{t} + \Fp = -(1-\beta)\C {\bf{X}}_{t},
\end{align}
where 
$\C=\mathcal{R}\hat{\mathbf{t}}\hat{\mathbf{t}} + \hat{\mathbf{n}}\hat{\mathbf{n}}$ is the drag tensor.  The term $\F(\X)$ includes the contributions from the follower force and the forces from bending and tension. Time is discretized using backward-Euler with a time step of $\Delta{t}=10^{-5}$. 

For the value of $\mathcal{R}$ we use the expression
\begin{equation}
 R = \frac{1}{2}\left(\frac{\log(2L/a)+1/2}{\log(2L/a)-1/2} \right)
\end{equation}
from \cite{johnson1979flagellar,cox1970motion} where $L$ is the wavelength and $a$ is the radius. Because we compare with data for \textit{Chlamydomonas reinhardtii} flagella, we take $L=10\mu{m}$ and $a=0.15\mu{m}$, which leads to $R\approx 0.6138$.

We discretize the filament into $N_{s}=100$ equal segments with endpoints $\X_{j}=\X(s_{j})$ for $j=0\ldots N_{s}$. For the clamped boundary condition we fix $\X_{0}=(0,0)$ and $\X_{1}=(\Delta{s},0)$. The follower force is applied at the tip as $-\sigma(\X_{N_{s}}-\X_{N_{s}-1})/\Delta{s}$. The internal forces are computed using a discretized variational principle which naturally captures the boundary condition at the free end. 

The forces from deformation are negative of the variational derivative of the mechanical energy 
\begin{equation}
\mathcal{E}=\frac{1}{2}\int_{0}^{1} \kappa^2 ds + \frac{1}{2} \int_{0}^{1} \hat{T}^2 ds, 
\end{equation}
where $\kappa$ is the curvature. The first term corresponds to the energy of bending, and the second term corresponds to the energy from the tension. In the spatially discrete system, expressions for the forces are computed by taking the variational derivative of a discrete energy. 

The discrete curvature at the interior points ($j\ne 0, N_{s}$) is 
\begin{equation}
\kappa_{j} = \Big(\frac{\hat{\bf{n}}_{j+1/2} + \hat{\bf{n}}_{j - 1/2}}{2}\Big) \cdot \Big(\frac{\hat{\bf{t}}_{j+1/2} - \hat{\bf{t}}_{j - 1/2}}{\Delta s}\Big),
\label{disc_curv:eq}
\end{equation}
where the discrete tangent vector is
\begin{equation}
\hat{\bf{t}}_{j+1/2} = \frac{{\bf{X}}_{j+1} - {\bf{X}}_{j}}{\Delta s},
\end{equation}
and the discrete normal is the $\pi/2$ rotation of the tangent. Equation \eqref{disc_curv:eq} represents a discrete version of $\kappa= \hat{\bf{n}}\cdot {\partial \hat{\bf{t}}}/{\partial s}$.
Using the orthogonality of the normal and tangent on a segment, equation \eqref{disc_curv:eq} can be simplified to 
\begin{equation}
\kappa_{j} = \frac{\hat{\bf{n}}_{j-1/2} \cdot \hat{\bf{t}}_{j+1/2} - \hat{\bf{n}}_{j+1/2}\cdot \hat{\bf{t}}_{j-1/2}}{2\Delta s}.
\end{equation}
In our discrete model, inextensibility is enforced approximately by penalizing extension and compression. The discrete tension is 
\begin{equation}
  \hat{T}_{j+1/2} = k_{s}\Big(\Big|\frac{{\bf{X}}_{j+1} - {\bf{X}}_{j}}{\Delta s}\Big| - 1\Big).
\end{equation}
We take $k_{s}=10^{4}$ which results in variations of length that are on the scale of $k_{s}^{-1}=10^{-4}$.

\clearpage


\section{Fitting for $\lambda,$ $\beta$ as a function of polymer concentration}
\label{app:fitting}

\begin{figure}[hb]
\centering
\includegraphics[width=\singlepanelwidth]{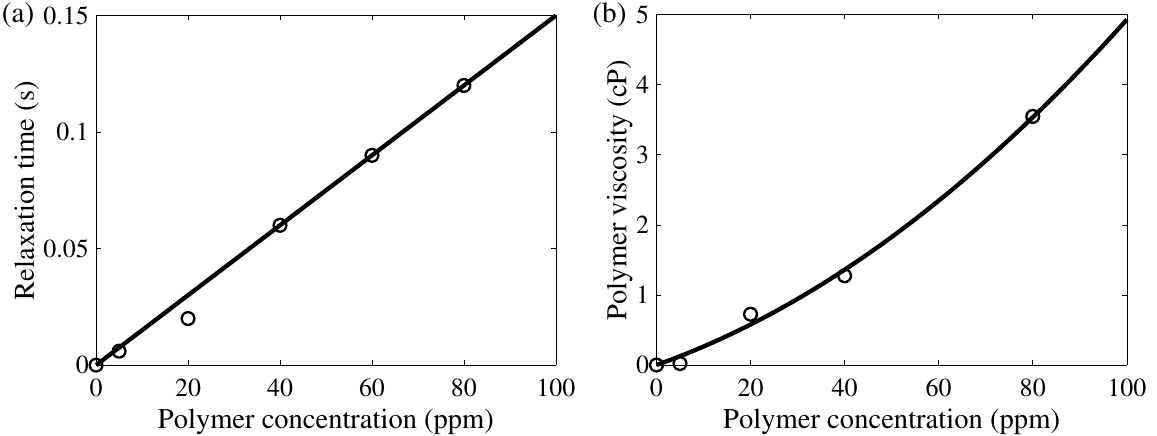}
\caption{Data and fit for relaxation time (a) and polymer viscosity (b) as a function of concentration of PAA. }\label{fig_fits}
\end{figure}

Here we describe how  we model  the dependence of the relaxation time and viscosity ratio  on concentration.  In \cite{qin2015flagellar} relaxation times for a range of PAA conectrations (from 5-80 ppm) are reported (see Table 1 in SI). We use a linear fit through the origin for this data and find a the dimensional relaxation time to be  $\tau = 0.0015 \c$ seconds, where $\c$ is the PAA concentration in ppm.  The data and our fit are plotted in Fig.\ \ref{fig_fits} (a). 

To model the dependence of the viscosity ratio on the polymer concentration we use the data reported in Fig.\ 2 of the SI in \cite{qin2015flagellar}. In this figure the authors plot the shear viscosity over a range of shear rates and find that they are nearly constant for the relevant range of shear rates.  In the case of the highest molecular weight in which shear thinning is observed, the  mean of the shear rate near the body ($15s^{-1}$) and the shear rate near the flagella ($50s^{-1})$ is used to estimate the relevant viscosity. We use a quadratic fit through the origin for the polymer viscosity and find  $\eta_p = .026\c+.00024\c^2.$ The data and our fit are plotted in Fig.\ \ref{fig_fits} (b).

\clearpage


\section{Varying Stiffness} 
\label{vary_stiffness:appendix}

Comparing the model predictions as functions of the dimensional relaxation time and viscosity requires choosing values for the parameter combinations $k_b/L^4$ and $\sigma/\sigma_0$. We used $k_b/L^4=0.25$ and $\sigma/\sigma_0=1.5$ in the simulations presented in Fig. \ref{fig_vcc}.  In order to explore how these choices affect our results, we consider a range of $k_b/L^4$ from $0.1 - 2$ and $\sigma/\sigma_0$ from $1.25 - 2.5.$  To explore parameters efficiently, we use the frequencies predicted from linear stability analysis rather than from the numerical simulations. These two different methods of computing the frequency agree well near the bifurcation but start to diverge away from the bifurcation. 

\begin{figure}[bht]
\centering
\includegraphics[width=\singlepanelwidth]{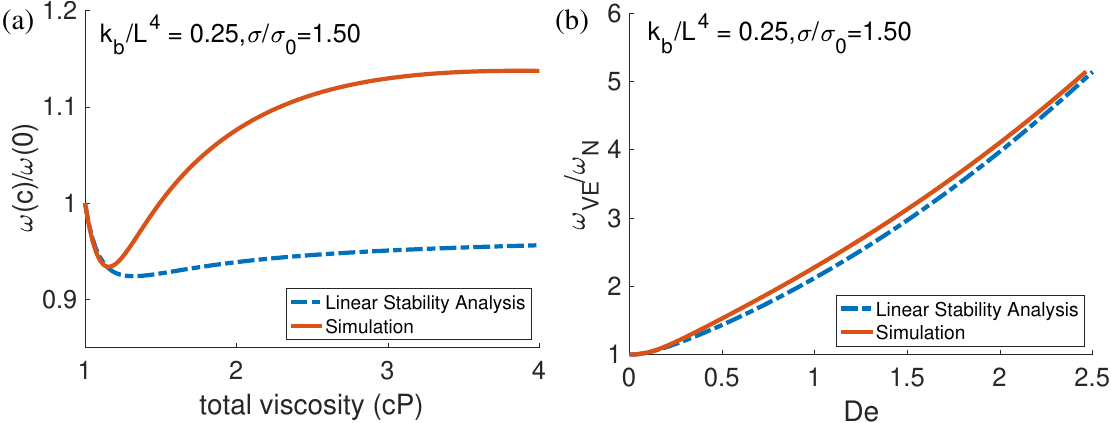}
\caption{(a) Simulations and LSA for frequency vs. total viscosity at 
parameters in main paper. (b) Simulations and LSA for normalized frequency vs. $\DE$ at 
parameters in main paper. }\label{fig_vcc_simlsa}
\end{figure}

Before we present the results for the wider range of parameters we show a comparison of the results from the simulations and the linear stability analysis for the values presented in the manuscript. Figure \ref{fig_vcc_simlsa}  compares the linear stability analysis and simulation computation of the frequency for the parameters $k_b/L^4=0.25$ and $\sigma/\sigma_0=1.5$ (presented in Fig. \ref{fig_vcc}).  In Fig.\ \ref{fig_vcc_simlsa} (a)  the simulations predict larger frequency at high polymer viscosity than the analysis. Nevertheless the qualitative feature of the non-monotonic dependence of frequency on viscosity in viscoelastic fluids is still predicted by the analysis.  Fig.\ \ref{fig_vcc_simlsa} (b) shows that the simulation and analysis show close agreement for the frequency boost seen in viscoelastic fluids over viscous fluids as the Deborah number is increased. 

\begin{figure}
\centering
\includegraphics[width=\singlepanelwidth]{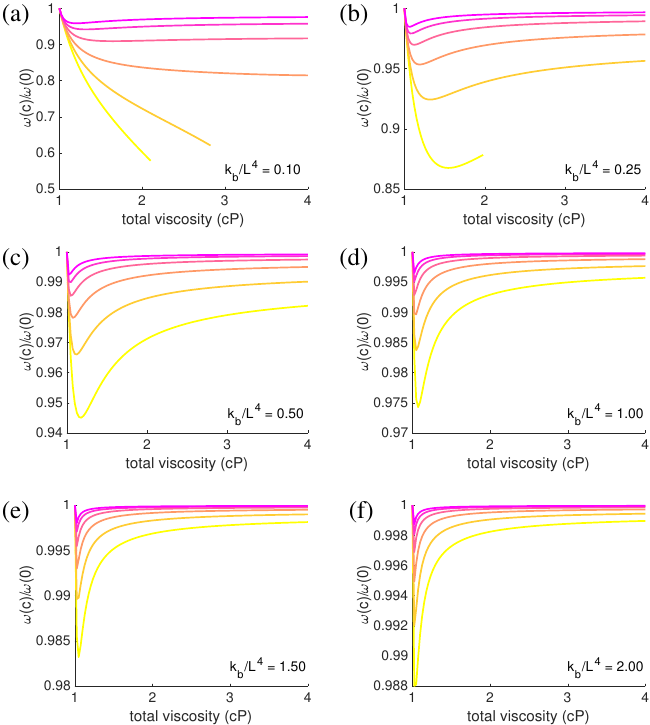}
\caption{Frequency scaled by its value at $c=0$ as a function of the total viscosity for a range of parameters: $0.1 \leq k_b/L^4 \leq 2$ and $1.25 \leq \sigma/\sigma_0 \leq 2.5.$ Color indicates value of $\sigma/\sigma_0=1.25,1.50,1.75,2.00,2.25,2.50$ ranging from light to dark. }\label{fig_vcc_lsa_vtv}
\end{figure}

\begin{figure}
\centering
\includegraphics[width=\singlepanelwidth]{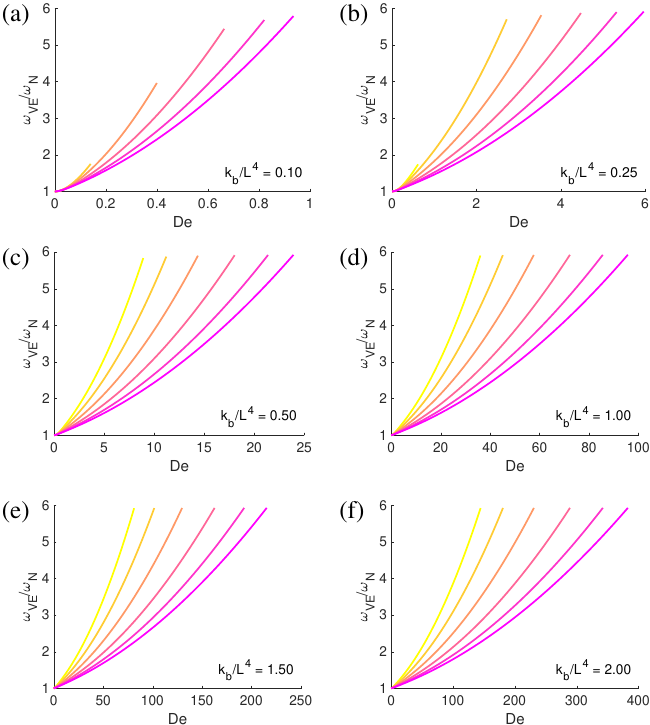}
\caption{Frequency in a viscoelastic fluid relative to frequency in a Newtonian fluid of the same viscosity as a function of $\DE$ for a range of parameters: $0.1\leq k_b/L^4 \leq 2$ and $1/25 \leq \sigma/\sigma_0 \leq 2.5.$ Color indicates value of $\sigma/\sigma_0=1.25,1.50,1.75,2.00,2.25,2.50$ ranging from light to dark.}\label{fig_vcc_lsa_vde}
\end{figure}

In Figs. \ref{fig_vcc_lsa_vtv} and \ref{fig_vcc_lsa_vde}  results of the exploration of the parameters for the ranges $0.1\leq k_b/L^4\leq 2$ and $1.25\leq \sigma/\sigma_0 \leq 2.5$ are presented using the linear stability analysis calculation for frequency.  In Fig.\ \ref{fig_vcc_lsa_vtv} we plot the normalized frequency $\omega(c)/\omega(0)$ as a function of the total viscosity.

Each panel in the figure shows the results for a single value of $k_b/L^4$ for increasing forcing strength from $1.25$ to $2.5$ corresponding to the colors going from light to dark. All but the smallest value of $k_b/L^4,\sigma/\sigma_0$ predict the non-monotonic response of frequency as a function of total viscosity.  The exact parameter values will change the amount that the frequency is predicted to decrease as well as where that predicted decrease is maximized, but a wide range of parameter values show the qualitative behavior reported in Fig. \ref{fig_vcc} (a).  

In Fig. \ref{fig_vcc_lsa_vde} the  frequency $\omega_{VE}/\omega_{N}$ is plotted as a function of $\DE$. As before each panel shows the reults different value of $k_b/L^4$ for a range of forcing strengths from $1.25$ to $2.5$ corresponding to the colors going from light to dark. The difference between these graphs is what values of $\DE$ are sampled for the given mechanical parameters. They all show the same qualitative behavior reported in Fig. \ref{fig_vcc} (b).

\clearpage


\section{Additional Viscoelastic Eigenvalue} 
\label{app:second_root}
The eigenvalues in a viscous fluid, $\etaV$, are related to those in a viscoelastic fluid, $\etaVE$ by the equation
\begin{equation}
  \etaV = \frac{(1-\beta)\etaVE} {\lambda\etaVE  + 1} + \beta \etaVE.
\end{equation}
This equation is equivalent to a quadratic equation in $\etaVE$, and so for each value fo $\etaV$ there are two values of $\etaVE$. In our numerical calucation of the viscoelastic eigenvalues, we identified the root with the largest real part. We checked that the other root always had negative real part by numerically computing it over a wide range of parameters. Here we give more information about the second root.  

\begin{figure}[bht]
	\includegraphics[width=0.75\textwidth]{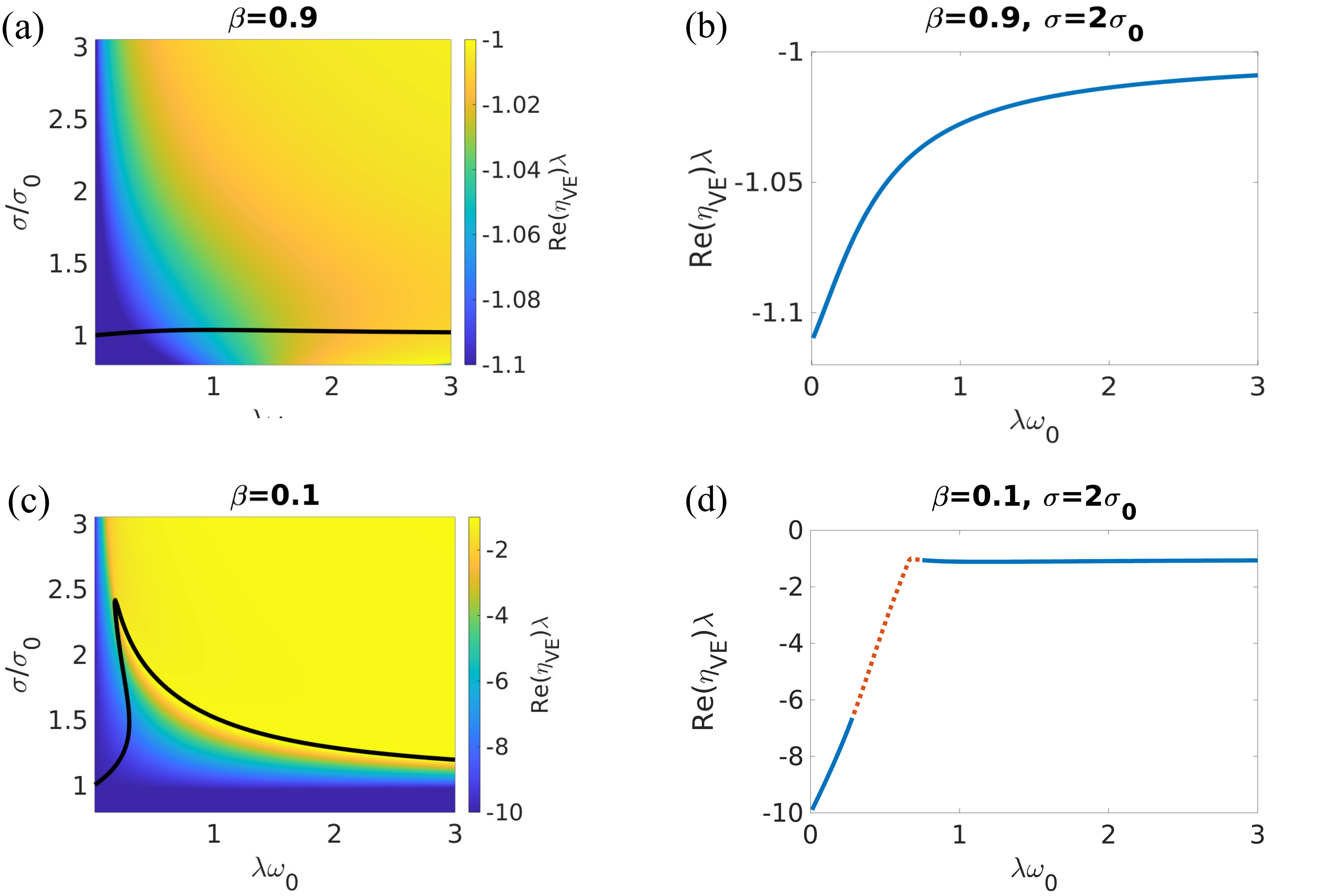}
  \caption{The product of the real part of the viscoelastic eigenvalue with smallest real part and the relaxation time for (a) $\beta=0.9$ and (c) $\beta=0.1$. The black line denotes the location of the bifurcation. This same quantity as a function of $\lambda\omega_{0}$ for the fixed value of $\sigma=2\sigma_{0}$ for (b) $\beta=0.9$ and (d) $\beta=0.1$. These plots help illustrate that $\etaVE\lambda \rightarrow -1/\beta$ as $\lambda\rightarrow0$ and $\etaVE\lambda \rightarrow -1$ as $\lambda\rightarrow\infty$. The red dotted line on (d) represents values in the stable region where both eigenvalues have negative real parts.} 
  \label{secondroot:fig} 
\end{figure}

We can obtain asymptotic expressions for both roots in the limits $\lambda\rightarrow 0$ and $\lambda\rightarrow\infty$. The obvious expressions for $\etaVE$ in these two limits are
\begin{align}
   & \etaVE = \etaV + \mathcal{O}(\lambda)  \text{ as } \lambda\rightarrow 0 \\
   & \etaVE = \frac{\etaV}{\beta} +  \mathcal{O}\left(\lambda^{-1}\right)   \text{ as } \lambda\rightarrow \infty.
\end{align}
The other roots in these limits are
\begin{align}
   & \etaVE = -\frac{-1}{\beta\lambda} + \mathcal{O}(1)  \text{ as } \lambda\rightarrow 0 \\
   & \etaVE = -\frac{1}{\lambda} +  \mathcal{O}\left(\lambda^{-2}\right)   \text{ as } \lambda\rightarrow \infty.
\end{align}
Note that in both of these limits this other eigenvalue always has negative real part, and is independent of $\etaV$. In both limits this other eigenvalue scales with $1/\lambda$, which suggests it is related to the fluid relaxation time scale. We observed that for all values of $\sigma$, $\lambda$, and $\beta$ considered in this paper the real part of this other eigenvalue was always less than $-1/\lambda$. In Figure \ref{secondroot:fig}, we show the product of the real part of this eigenvalue  with $\lambda$ for $\beta = 0.9$ and $\beta=0.1$ in the $\lambda\omega_{0}$\---$\sigma/\sigma_{0}$ plane.  To help illustrate the asymptotic results, we display the product of the real part of this eigenvalue  with $\lambda$ for fixed $\sigma=2\sigma_{0}$ as a function of $\lambda$. 

\clearpage


\section{Comparing Asymptotic and Numerical Solutions}
\label{app:asym_numer_comp}
\begin{figure}[hb]
\centering
\includegraphics[width=0.3\textwidth]{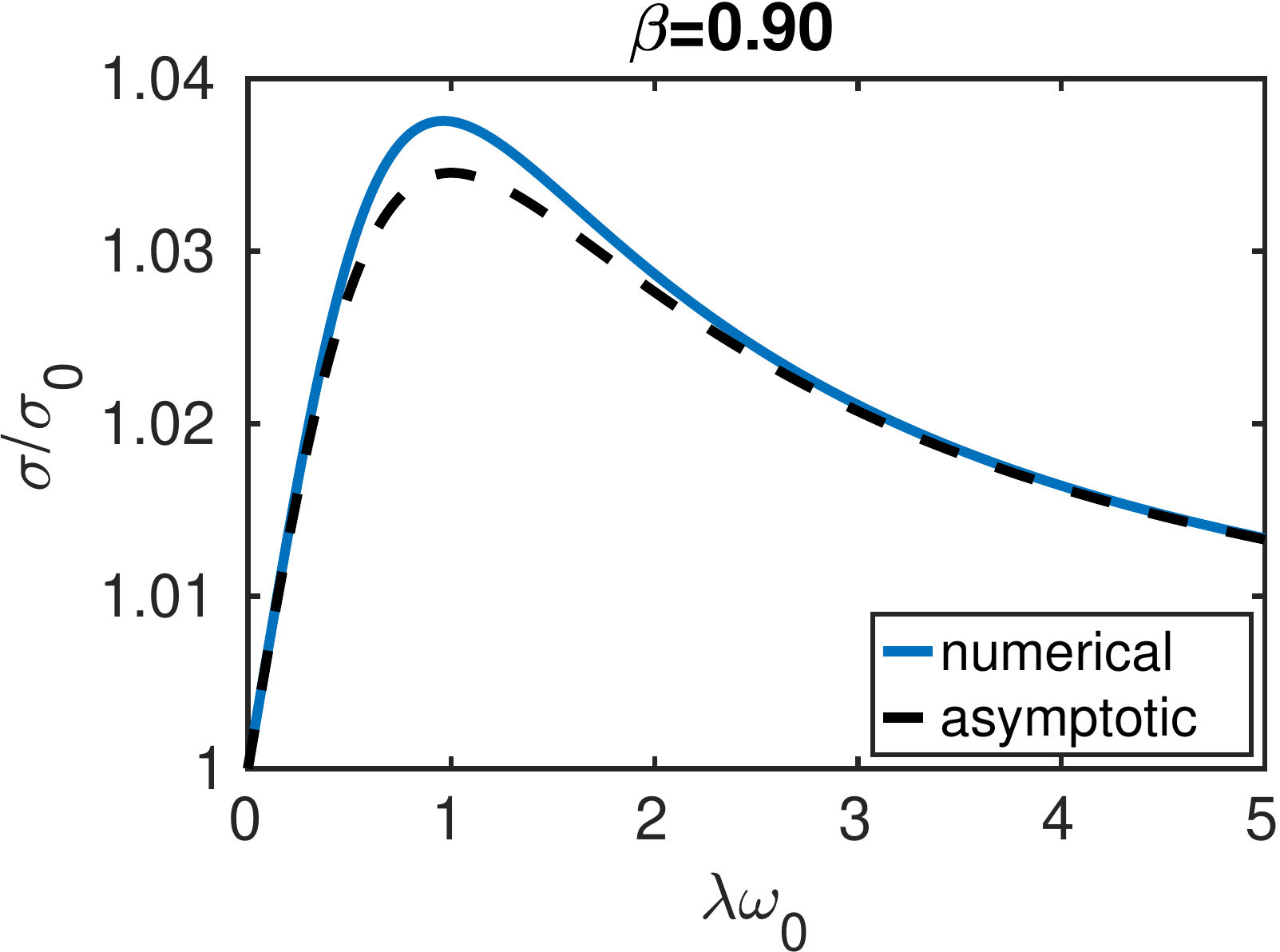}
\includegraphics[width=0.3\textwidth]{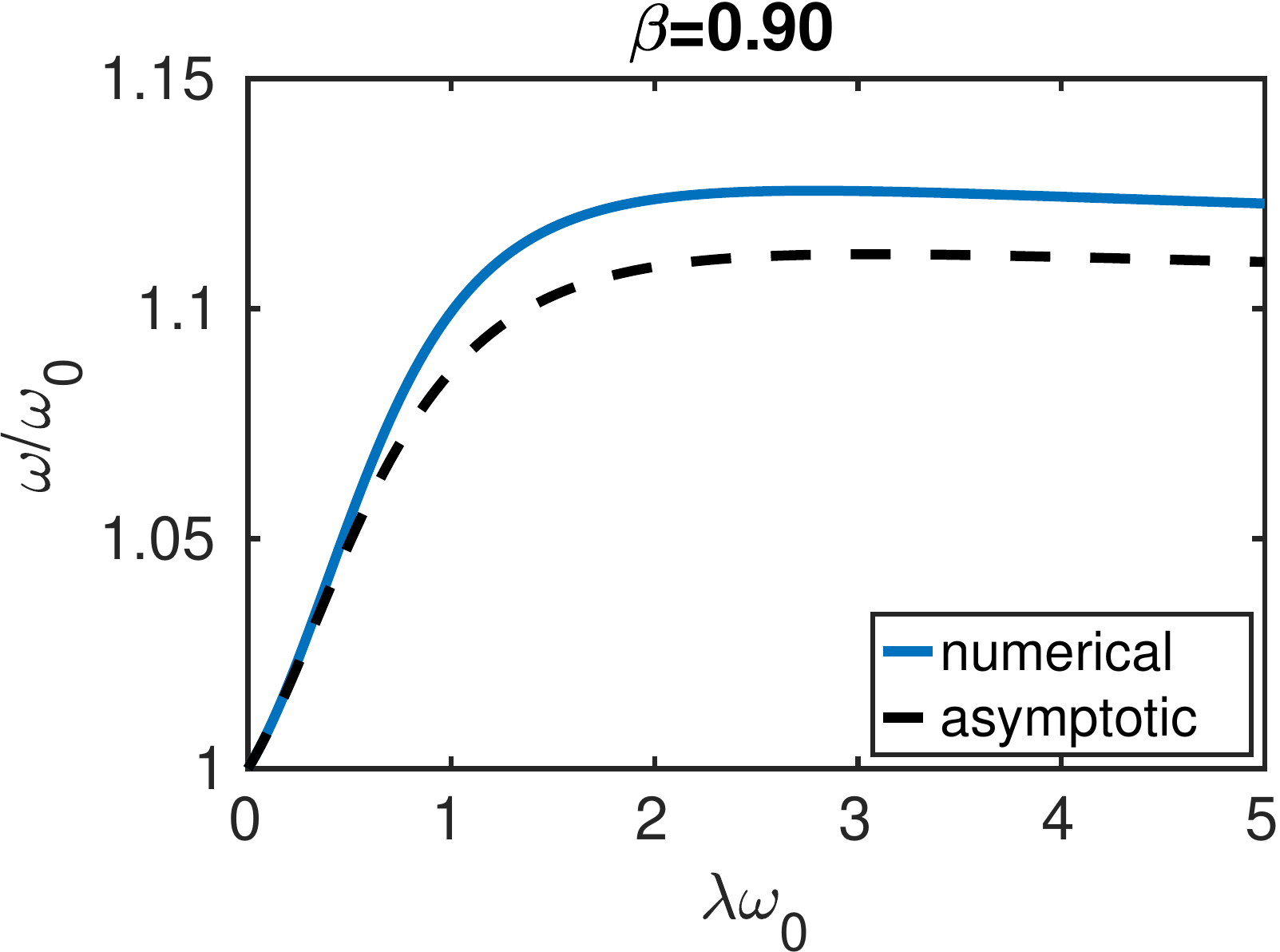} \\[10pt]
\includegraphics[width=0.3\textwidth]{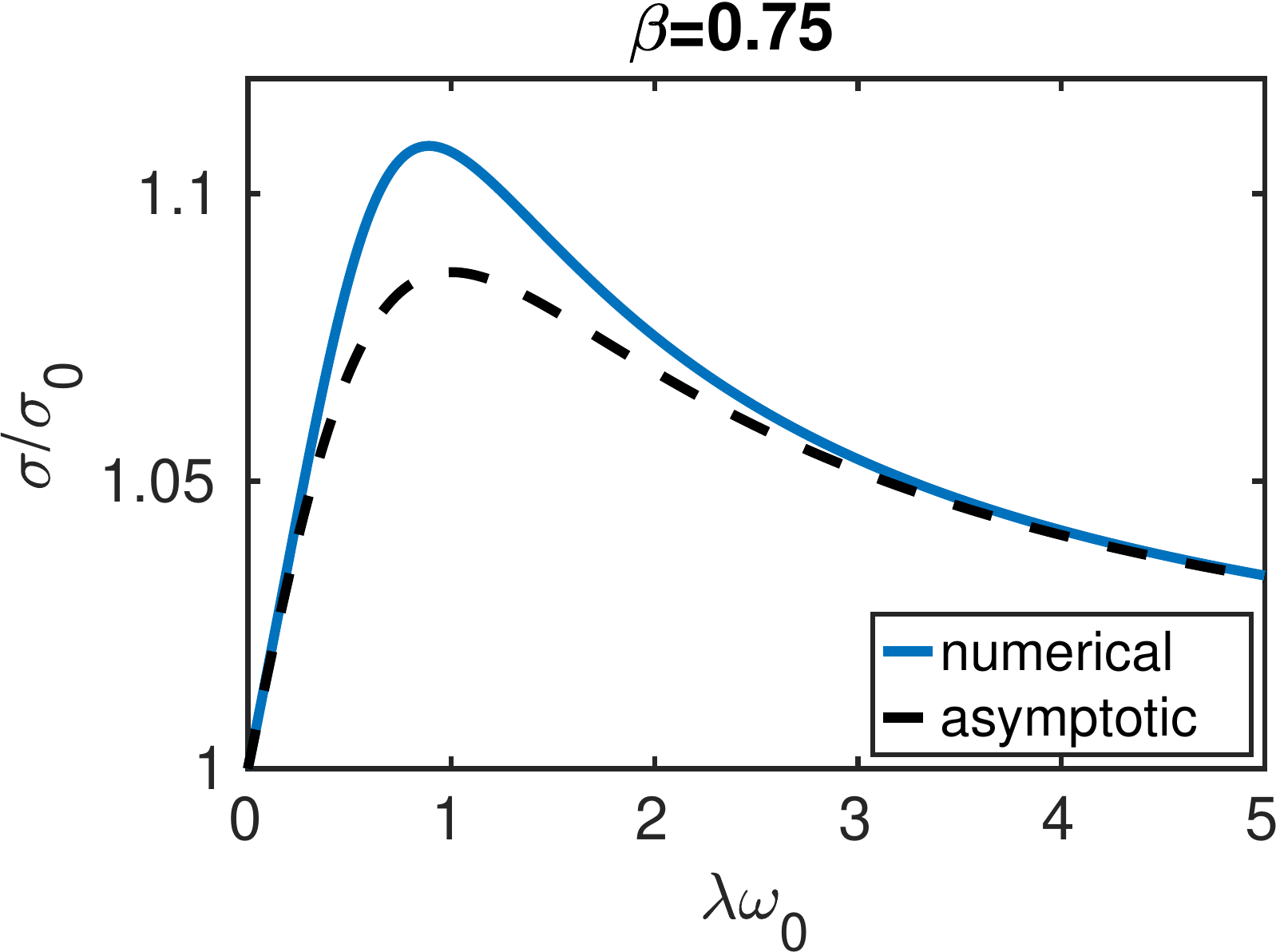}
\includegraphics[width=0.3\textwidth]{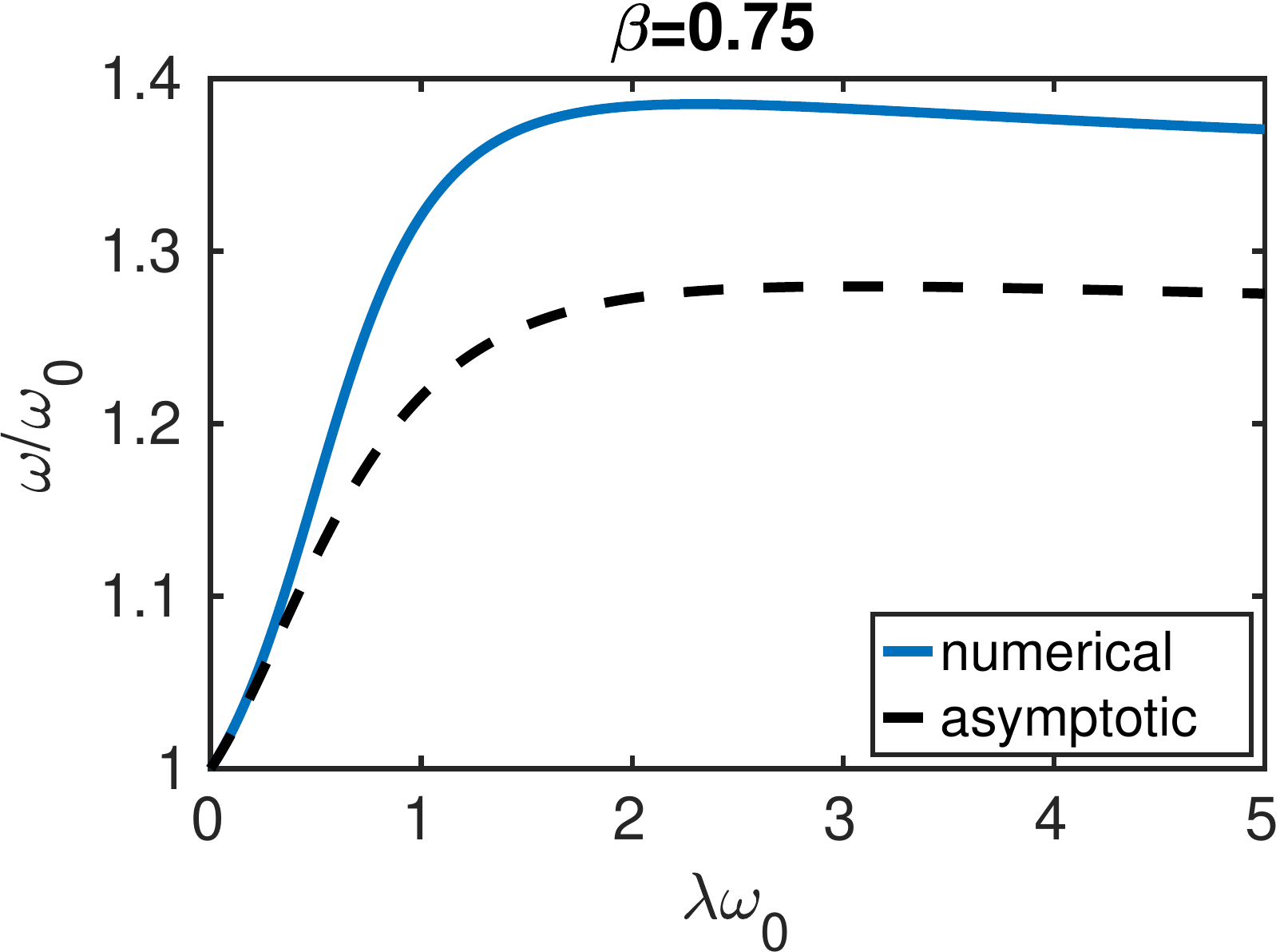} \\[10pt]
\includegraphics[width=0.3\textwidth]{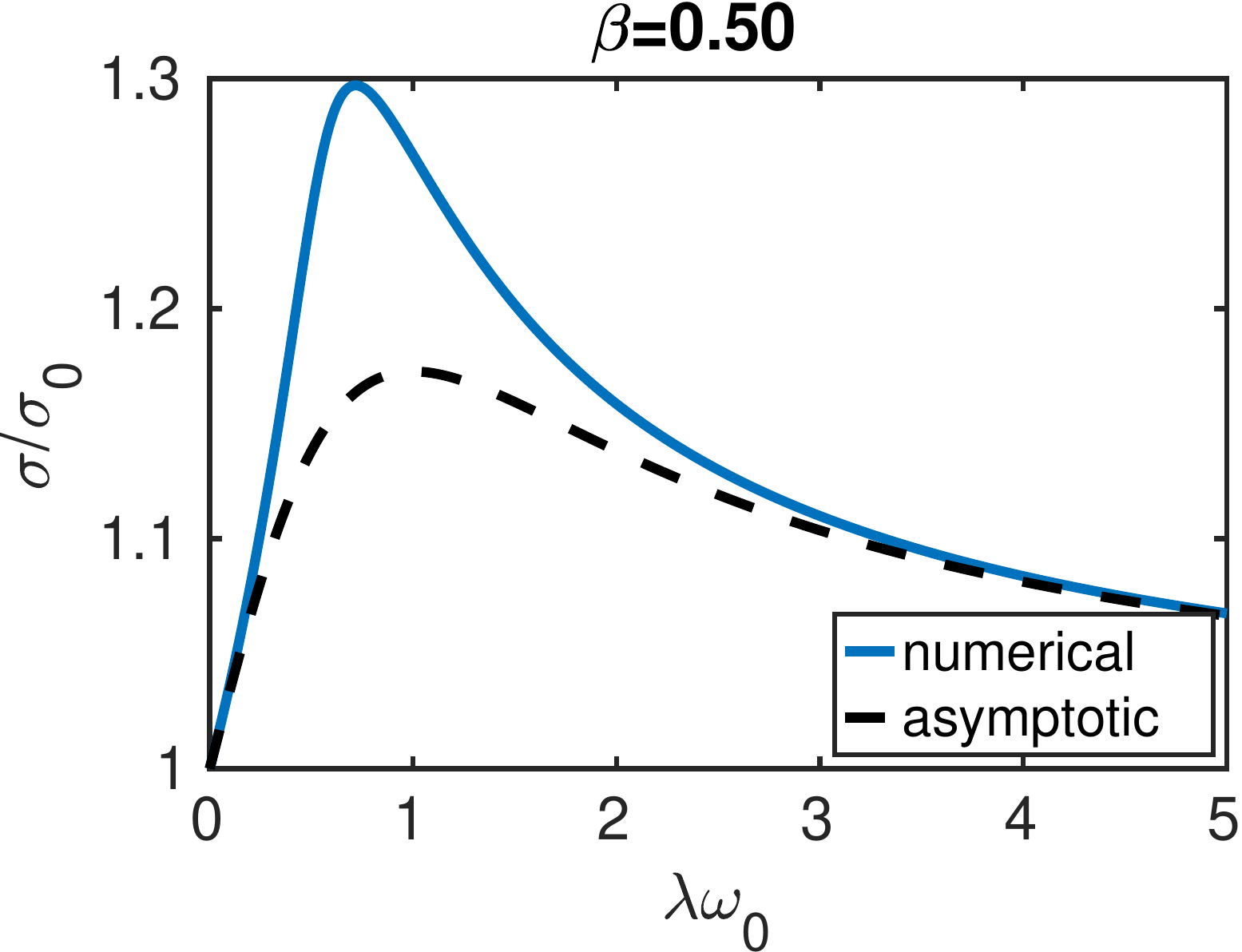}
\includegraphics[width=0.3\textwidth]{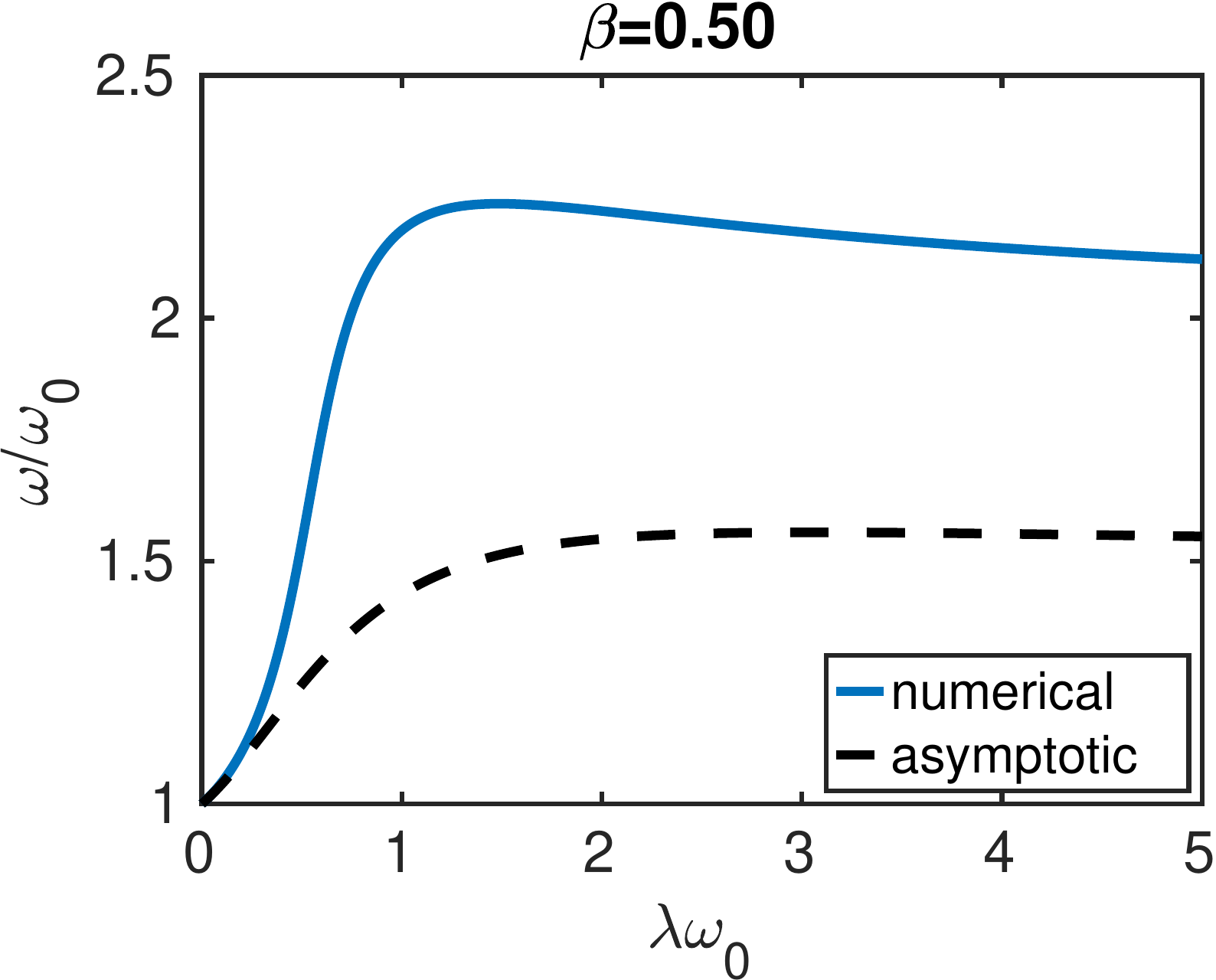} \\[10pt]
\includegraphics[width=0.3\textwidth]{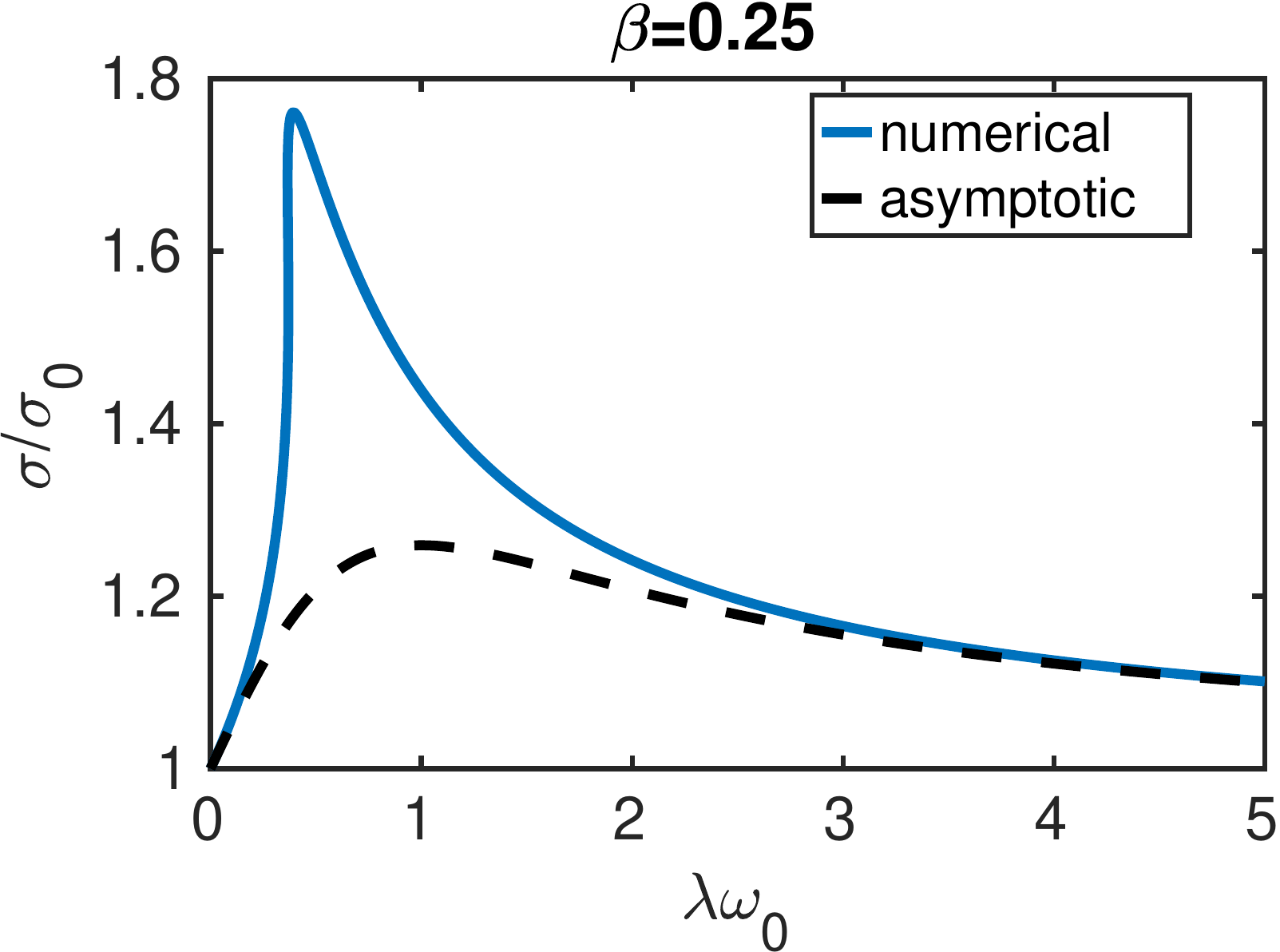}
\includegraphics[width=0.3\textwidth]{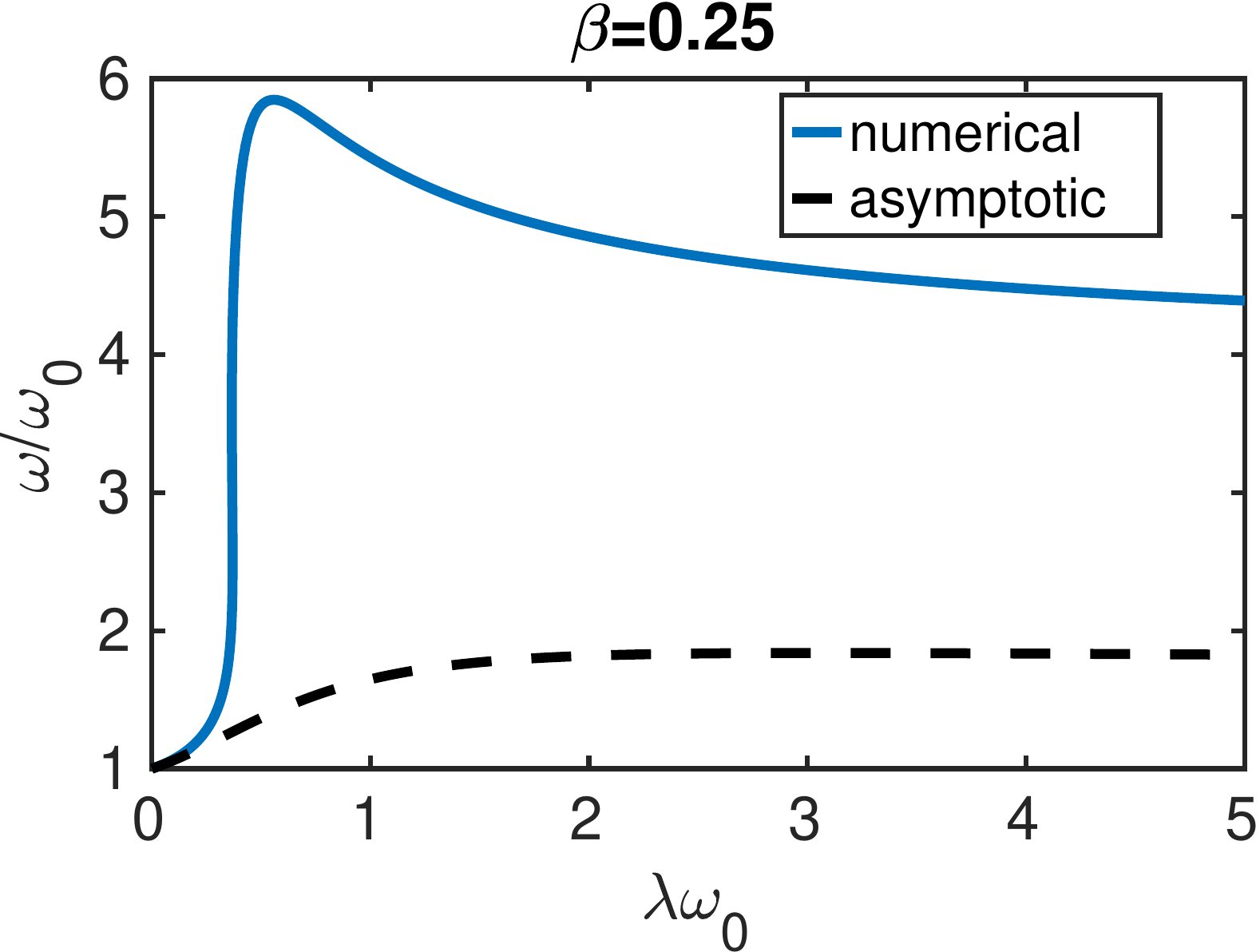} \\[10pt]
\caption{Plots of the asymptotic and numerical solutions for the critical follower force (left column) and emergent frequency (right column) at the bifurcation as functions of the relaxation time for different values of $\beta$. The asymptotic expressions for the critical force and frequency are given in equations \eqref{sigma_asym_small_beta:eq} and \eqref{omega_asym_small_beta:eq}, respectively.}
\label{asym_numer_comp:fig}
\end{figure}

\clearpage


\section{Comparison with a Motor Model}
\label{app:motorSI}
%
%
%
%
%

Here we show that for an active filament driven by the model for molecular motor activity from Camalet and Julicher \cite{Camalet_2000}, in the limit of vanishing polymer viscosity, the expression for the frequency as a function of relaxation time is of the same form as that for a filament diven by a follower force  given in \eqref{omega_asym_small_beta:eq}. Thus it is expected that the frequency generally increases with relaxation time and approaches a constant value in the limit of large relaxation time. These two features of the frequency response are consistent with the experimental data from \cite{qin2015flagellar}.

The structure of the flagellar axoneme consists of an arrangement of microtubule doublets  containing dynein crossbridges whose action generates sliding forces between the doublets. We consider a commonly used two-dimensional model to describe planar beating that reflects the structure of pairs of microtubule filaments connected by molecular motors.  Consider the filament to be composed of two parallel inextensible filaments separated by a fixed distance $a$. At the base the filaments are both clamped. Molecular motors generate equal and opposite forces tangental to each filament. If the filaments were free at both ends, these forces would cause the filaments to slide past one another. Because both filaments are clamped at the base, the motor activity generates active moments that drive bending of the filaments. The linearized equation for the vertical displacement is 
\begin{equation}
  -\xi_{\perp} y_{t}  -k_{b}y_{ssss} + a f_{s} =0,
  \label{beam_with_motor:eq} 
\end{equation}
where $f=f(s,t)$ represents the motor force density along the filament. 

Closing the system requires a model for the motor force density; see, for example, \cite{Riedel-Kruse:2007:HSFP:motorsshape,gallagher2023axonemal,oriola2017nonlinear}. Another approach involves examining the linearized equations near the bifurcation point (i.e.\ assuming the filament is undergoing small amplitude periodic oscillations) where it is assumed that the force is linearly related to the shape of the filament \cite{Camalet_2000,Sartori:2016:Elife:dynamiccurvature}. Following  \cite{Camalet_2000}, they assume that $f = \hat{f}\exp(i\omega t) + \bar{\hat{f}}\exp(-i\omega t)$ and 
\begin{equation}
  \hat{f} = \chi(\sigma,\omega) \hat{\Delta} + \mathcal{O}(\hat{\Delta}^3)
  \label{f_chiDelta:eq}
\end{equation}
where $\Delta$ is the sliding displacement between the pair of filaments, which is related to the vertical displacement by 
\begin{equation}
  \Delta = a y_{s} + \mathcal{O}(y^3). 
  \label{Delta_def:eq}
\end{equation}
The function $\chi$ is the linear response function, which depends on the motor model, but is assumed to depend on the frequency $\omega$ and a control parameter $\sigma$.

Putting together \eqref{beam_with_motor:eq}\--\eqref{Delta_def:eq} and appropriately nondimensionalizing results in   
 \begin{equation}
    -\hat{y}_{ssss} + \chi \hat{y}_{ss} =i\omega \hat{y},
\end{equation}
where we are using the convention that $\chi$, $\omega$, and $\sigma$ represent dimensionless quantities. Notice that if $\chi=-\sigma_{0}$ and $\etaV=i\omega$, this equation is of the same form as that obtained for a filament subject to a follower force at the bifurcation point in a viscous fluid; i.e.\ compare with equations \eqref{eig_val1:eq}\--\eqref{eq:ve_eig_val} with $\beta=1$. A key difference is that the function $\chi$ is not a constant, and in particular, it depends on the frequency of the beat.  

We can thus reexamine the asymptotic analysis in the limit $\beta\rightarrow 1$ for this motor model in place of the follower force model, and we consider the case where the eigenvalues depend additionally on the frequency. In the analysis of Section \ref{betato1:sec}, we linearize the dependence of the viscous eigenvalue on the follower strength, which plays the role of the ``control parameter'' in the follower force problem. We take this same approach except that the eigenvalue, $\etaV$, will depend on both the control parameter, $\sigma$, and the frequency, $\omega$. Suppose that the bifurcation point occurs at $\sigma_{0}$ with frequency $\omega_{0}$. Near the bifurcation point
\begin{equation}
  \etaV = i\omega_{0} + \hat{\sigma}\frac{\partial\etaV}{\partial \sigma} + \hat{\omega}\frac{\partial\etaV}{\partial \omega} + h.o.t.,
  \label{etaV_linearize_motor}
\end{equation}
where $\hat{\sigma}=\sigma-\sigma_{0}$ and $\hat{\omega} = \omega-\omega_{0}$. 

We now look for the viscoealstic bifurcation point when $1-\beta=\epsilon$ is small. As in the follower force problem, the viscous and viscoelastic eigenvalues are related by
\begin{equation}
  \etaV\left(\sigma,\omega\right) = \frac{(1-\beta)i\omega} {\lambda i\omega  + 1} + \beta i\omega.
  \label{eq:ve_eig_at_bif}
\end{equation}
Linearize the left side of this equation using \eqref{etaV_linearize_motor} and equate real and imaginary parts to obtain
\begin{align}
   a_{\sigma}\hat{\sigma} + a_{\omega}\hat{\omega}& = \frac{\epsilon \omega^{2}\lambda}{1 + \omega^{2} \lambda^{2}} 
                     + h.o.t.,\\
   \omega_{0} + b_{\sigma}\hat{\sigma} +b_{\omega}\hat{\omega} &= 
        \omega \bigg(1 - \epsilon \frac{\omega^{2}\lambda^{2}}{1 + \omega^{2}\lambda^{2}}\bigg)
        + h.o.t.,
\end{align}
 where ${\partial\etaV}/{\partial \sigma} = a_{\sigma}+b_{\sigma}i$ and ${\partial\etaV}/{\partial \omega} = a_{\omega}+b_{\omega}i$. Solve the first equation for $\hat{\sigma}$ and plug in to the second equation to eliminate $\hat{\sigma}$. Finally expand $\omega=\omega_{0} + \epsilon\omega_{1} + \mathcal{O}(\epsilon^2)$ and $\hat{\omega} = \epsilon\omega_{1}+ \mathcal{O}(\epsilon^2)$ to arrive at the expression for the frequency
\begin{equation}
\omega  = \omega_{0} + (1-\beta)\omega_{0}\left(\frac{\frac{b_{\sigma}}{a_{\sigma}}\lambda \omega_{0} + \lambda^{2}\omega_{0}^2}{1 + \lambda^{2}\omega_{0}^2}\right)\left(1-z\right)^{-1} + \mathcal{O}\bigl((1-\beta)^2\bigr),
\end{equation}
where 
\begin{equation}
  z = b_{\omega}-\frac{a_{\omega}b_{\sigma}}{a_{\sigma}}.
\end{equation}
Notice that $z$ contains all the dependence of the motor activity on the frequency. When $z=0$, we recover the result in the paper for the follower force.

This expression for the frequency is of the same form as that obtained for the filament driven by a follower force given in equation \eqref{omega_asym_small_beta:eq}. Thus this analysis predicts that if $z<1$, then in the limit of small polymer viscosity a filament driven with internal motors consistent with \eqref{f_chiDelta:eq} will exhibit a frequency that initially increases with relaxation time and approaches and a constant in the limit of large relaxation time. These two features of the frequency response are consistent with follower model and the experimental data from \cite{qin2015flagellar}. 

\clearpage

\bibliographystyle{rspublicnat}
\bibliography{follower_citations,more_follower_citations}

\end{document}